\documentclass[aps,pop,showpacs,twocolumn,superscriptaddress,letterpaper]{revtex4-1}
\usepackage{mathrsfs}
\usepackage{amsmath}
\usepackage{bm}
\usepackage{color}
\usepackage{graphicx}
\usepackage{hyperref}
\usepackage{float}
\hypersetup{pdfpagemode=UseNone,pdfstartview=FitH,pdfstartpage=1}

\begin{document}


\title{Plasma Waves Accessibility Diagrams: A Tutorial to Include the Fluid and Kinetic Thermal Effects}
\author{Huasheng Xie}
\email[]{Email: huashengxie@gmail.com} \affiliation{Hebei Key Laboratory of Compact Fusion, Langfang 065001, China}
\affiliation{ENN Science and Technology Development Co., Ltd., Langfang 065001, China}
\author{Haojie Ma}\affiliation{Hebei Key Laboratory of Compact Fusion, Langfang 065001, China}
\affiliation{ENN Science and Technology Development Co., Ltd., Langfang 065001, China}
\author{Yukun Bai}\affiliation{Hebei Key Laboratory of Compact Fusion, Langfang 065001, China}
\affiliation{ENN Science and Technology Development Co., Ltd., Langfang 065001, China}
\date{\today}

\begin{abstract}
Although the accurate description of the wave propagation and absorption in plasmas requires complicated full wave solutions or kinetic simulations, the local dispersion analysis can still be helpful to capture the main physics of wave properties. Plasma wave accessibility informs that whether a wave can propagate to a region, which usually depends on the wave frequency, wave vector, and the local plasma density and magnetic field. In this tutorial paper, we describe the wave accessibility beyond usual textbooks and especially highlight the warm plasma effects. Useful numerical models and methods are provided, which can be used to obtain a quick view of the wave accessibility parameters space. The thermal effects are modeled by both multi-fluid model with isotropic pressure term and kinetic model with Maxwellian velocity distribution function. All cold plasma waves from high frequency electron cyclotron waves, intermediate frequency low hybrid waves to low frequency ion cyclotron waves, as well as kinetic ion and electron Bernstein waves, are presented. The questions that how many plasma wave modes exist and how to find the solutions are also discussed. It is interesting to find that the warm multi-fluid model, though incapable of reproducing the Bernstein modes, can provide a quick way to determine whether the thermal effects are important. To show the kinetic thermal effects, the ray tracing calculations of the mode conversion from cold plasma waves to kinetic waves are also provided, i.e., from the slow X electron cyclotron wave to electron Bernstein wave and from the ion cyclotron fast wave to ion Bernstein wave.
\end{abstract}


\maketitle

\section{Introduction and motivation}\label{sec:intro}

The existence of numerous waves and instabilities is one of the most important features of plasmas, which are usually be described by single fluid magnetohydrodynamics (MHD) model, multi-fluid cold and warm models, and kinetic model.  Due to the collisionless damping and collision effects, the plasma waves are found to be important approaches to heating plasma and driven current in magnetized confinement fusion research\cite{Freidberg2007,Cairns1991,Wesson2011}. To accurately study the wave propagation and absorption in plasmas requires complicated linear and quasi-linear full wave solution or nonlinear kinetic simulation. However, it is found that the linear theory can describe most of the physics properties of plasma waves. In study of plasma wave propagation and heating, local linear dispersion analysis is usually the first and even the most important step.

\begin{figure}[htbp]
\centering
\includegraphics[width=8cm]{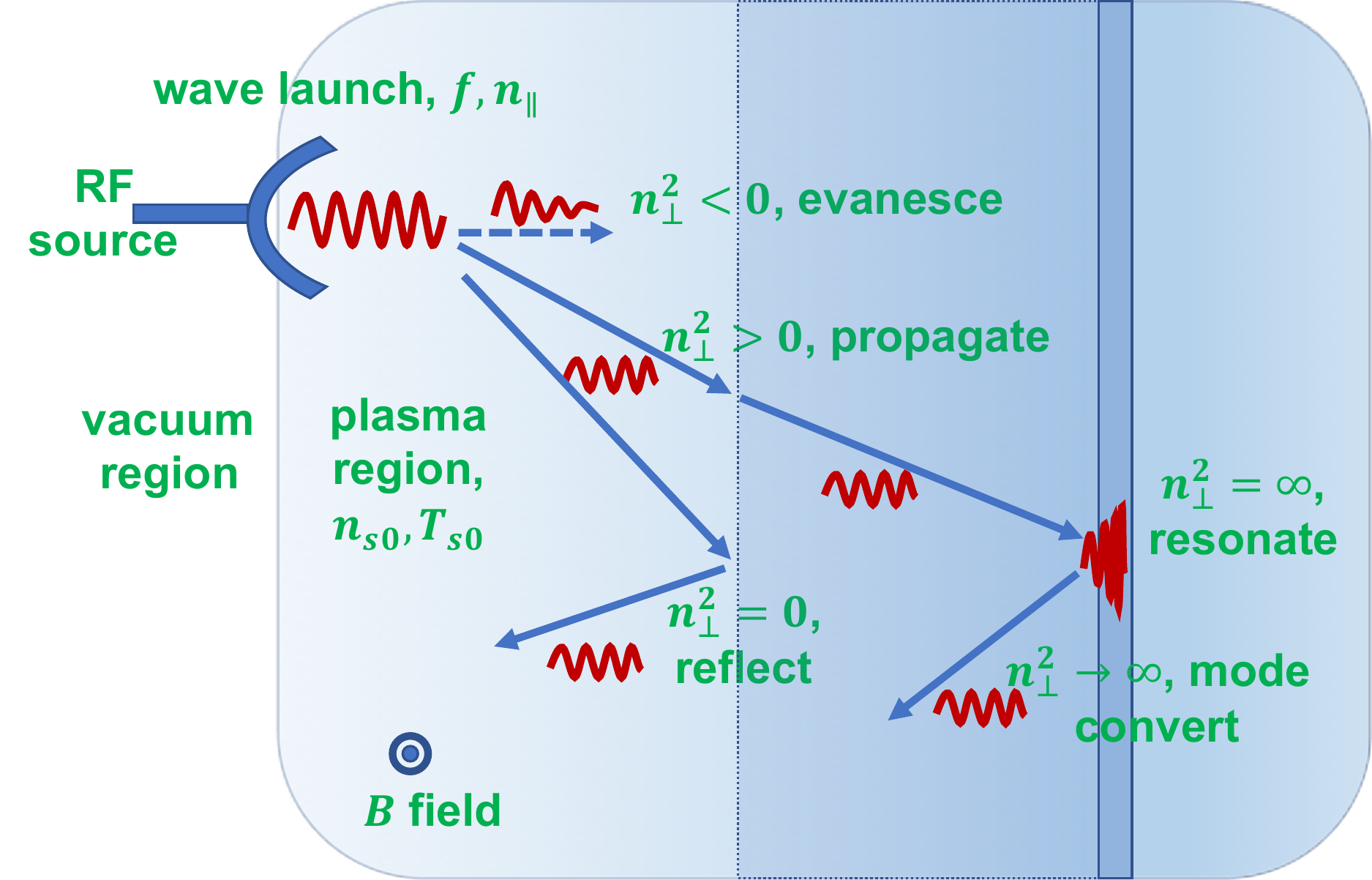}
\caption{A cartoon to show the plasma wave propagation, which represents the wave accessibility. The wave is launched at the outside region or plasma edge via the antenna with the frequency be determined by the radio frequency (RF) source.}\label{fig:bon_cartoon}
\end{figure}

In usual textbooks or monographes\cite{Stix1992,Swanson2003,Gurnett2005}, the local linear cold plasma theory is used to give the basic knowledges of a plasma wave based on figures such as $\omega$ v.s. $\bm k$, $n_\perp^2$ or $n^2$ vs. $\omega$, and  the Clemmow-Mullaly-Allis (CMA) diagram. Here, $\omega$ is the wave frequency, $\bm k$ is the wave vector,  $n$ ($n_\perp$) is the (perpendicular) refractive index. With the use of these figures, the wave accessibility under given plasma parameters can be readily identified. Local cold plasma models are also widely used to study the wave propagation in nonuniform plasma via WKB (Wentzel-Kramers-Brillouin) ray tracing method\cite{Batchelor1982,Smirnov2003,Peysson2012,Xie2021}, though the wave absorption or heating requires kinetic models.
In experimental wave heating studies, the  wave is launched from the outside vacuum region or in low density edge region with a fixed frequency $f=\omega/2\pi$ and some ranges of parallel wave vector $k_\parallel$. In the present tutorial, we mainly discuss waves in axisymmetric devices, i.e., with all profiles $f_0(r,\phi,z)$; such as magnetic field, plasma densities and temperatures have $\partial f_0/\partial\phi=0$ in cylindrical coordinates $(r,\phi,z)$. In tokamak\cite{Wesson2011,Freidberg2007}, $k_\parallel\sim k_\phi$, and $n_\phi=k_\phi r$ is conserved, which can be seen clearly from later Eq.(\ref{eq:dray}). In mirrors, field-reversed configurations (FRC) or dipoles, $k_\parallel\sim k_z$; usually $\partial f_0/\partial z\ll\partial f_0/\partial r$, i.e., $k_z$ changes much slower than $k_r$. Thus, it is a good approximation, at least at the wave launch region, to take $k_\parallel$ or $n_\parallel=k_\parallel c/\omega$ as a fixed input parameter, while calculating $k_\perp$ or $n_\perp=k_\perp c/\omega$. The $k_\parallel$ spectrum is usually determined by the antenna. Hence, in the wave accessibility study, we usually study the change of $n_\perp^2$ as a function of other parameters. If the real part of $n_\perp^2$, i.e., ${\rm Re}(n_\perp^2)>0$, the wave can propagate; if ${\rm Re}(n_\perp^2)<0$, the wave is evanescent; if ${\rm Re}(n_\perp^2)=0$, the wave cuts off and reflects back; if ${\rm Re}(n_\perp^2)\to\infty$, the wave is resonant at that position, in which case, mode conversion from one wave to another wave is possible. These are shown in a cartoon at Fig.\ref{fig:bon_cartoon}.

The following questions arise: How valid is the cold plasma approximation for the study of the wave accessibility and propagation? What is the best way to obtain a quick quantitative knowledge of wave accessibility, especially when  thermal effects are important?

To answer the above questions, we have developed a warm multi-fluid eigenvalue model which can yield all wave mode solutions. A series of wave accessibility diagrams are provided, via scanning $n_\perp^2$ v.s. frequency $f$, density $n_{s0}$, parallel refractive index $n_\parallel$, and also major radius $R$. These diagrams can be simpler than ray tracing to illustrate the basic wave propagation properties. They are also more useful than CMA diagram, since
the CMA diagram, showing wave propagation properties in plasma density and magnetic field, i.e., in a two parameters space, is limited to cold plasma, usually with only two-species, i.e., one electron and one ion. 

Besides the warm multi-fluid model, we also emphasize the kinetic effects. A non-relativistic and collisionless kinetic model is used to calculate the complex solutions of $k_\perp$. The imaginary part of $k_\perp$, i.e., ${\rm Im}(k_\perp)$ provides the information of wave heating/absorption. In kinetic model, we show that there possibly exist infinite wave mode solutions. A Cauthy contour method is used to locate all the solutions of $k_\perp$ in a chosen complex plane region. We conclude that, besides the multi-fluid modes, with kinetic corrections, the main important kinetic modes are electron and ion Bernstein modes. Other modes are usually strongly damped and thus are less physically interesting. Ray tracing examples of mode conversion from cold plasma waves to kinetic waves are also shown.

We assume that the readers have the basic knowledge of cold plasma waves.  The present tutorial can also be seen as an overview of basic pictures of plasma wave heating physics. In the following sections, Sec. \ref{sec:picwave}  discusses some basic pictures of plasma waves by the comparison of fluid and kinetic dispersion relation solutions. Sec. \ref{sec:model} gives the model equations used in the present tutorial. Sec. \ref{sec:accessibility} provides series wave accessibility diagrams. Sec. \ref{sec:ray} shows some ray tracing results. Appendix \ref{sec:howmanymodes} discusses how many kinetic wave modes exist. Sec. \ref{sec:summ} summarizes the present study.






\begin{widetext}

\begin{figure}
\centering
\includegraphics[width=14cm]{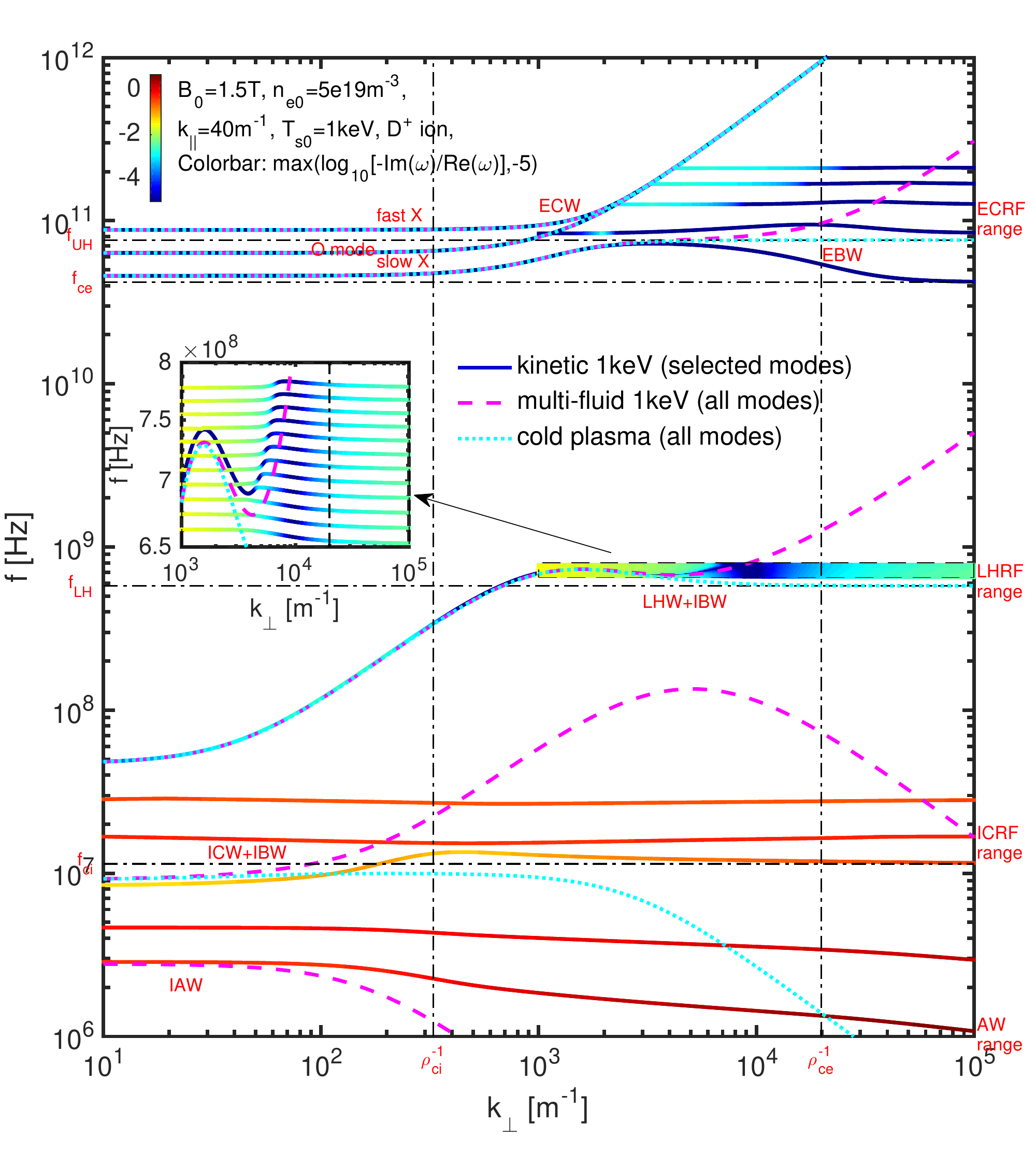}
\caption{Typical plasma waves in a fusion device. Parameters are the same to Ref.\cite{Pinsker2015} Fig.7, except that we have added the temperature effects with $T_i=T_e=1$keV.}\label{fig:pinsker15_fig7}
\end{figure}

\end{widetext}

\section{Basic pictures of plasma waves}\label{sec:picwave}

Basic properties of plasma waves are usually studied via dispersion relations, the model equations and detailed notations of which will be given in Sec. \ref{sec:model}.  Several excellent textbooks or monographes, c.f.\cite{Stix1992,Swanson2003,Gurnett2005}, have well described the plasma waves. For example, five branches of cold plasma waves are discussed in detail, which are found to be very useful to understand most of the wave features. Swanson\cite{Swanson2003} gives typical wave figures for overdense ($\omega_{pe}>|\omega_{ce}|$) and underdense ($\omega_{pe}<|\omega_{ce}|$) plasmas, and also discuss the multi-fluid plasma waves. The kinetic electron and ion Bernstein waves are also discussed by them. Before reading the following part of the present tutorial, we recommend that the readers read some of these textbooks to have a basic knowledge of plasma waves, i.e., ion and electron cyclotron waves, low and high hybrid waves; Bernstein waves; etc.

Here, we use one $\omega$ v.s. $k_\perp$ figure,  Fig.\ref{fig:pinsker15_fig7}, to show the basic pictures of cold, warm multi-fluids and kinetic plasma waves, which also demonstrates the thermal effects well. Typical realistic tokamak plasma parameters, magnetic field $B_0=1.5{\rm T}$, electron density $n_{e0}=5\times10^{19}{\rm m}^{-3}$ and deuterium ion, are chosen to provide the numerical solutions of corresponding dispersion relations in Sec. \ref{sec:model}. We also have fixed $k_\parallel=40{\rm m}^{-1}$, and set temperatures $T_{s0}=1{\rm keV}$ to study the thermal effects. These parameters are taken the same as those in Fig.7 of Ref.\cite{Pinsker2015}, except that we add the temperature $T_{s0}$.

Firstly, the cyan dot lines show all the five branches of cold plasma waves from high frequency to low frequency, with $T_{s0}=0$. In laboratory plasma wave heating studies, these waves are usually distinguished by the frequency range. The two highest frequency branches ($\sim100{\rm GHz}$) are named as electron cyclotron wave (ECW), since which are in the electron cyclotron range of frequency (ECRF) with $f\sim f_{ce}$. They are also called as electron cyclotron (fast) extraordinary mode (X-mode) and ordinary mode (O-mode), which are actually the two vacuum electromagnetic waves $\omega^2\sim k^2c^2$ modified by plasmas. That is, these two branches are the only two branches which can be launched from the outside of plasma and can propagate in vacuum. There is no dependence on the magnetic field strength for O-mode, with $\omega^2\to\omega_{pe}^2$ for $k\to 0$; whereas the mode of propagation depends on the magnetic field strength for X-mode. There is another X-mode with frequency slightly below the O-mode frequency, which is usually called as electron cyclotron slow X-mode. Here, `slow' means the phase velocity $v_p=\omega/k$ is smaller than the fast one. This slow X-mode does not exist in vacuum. At the intermediate frequency  ($\sim1{\rm GHz}$), we see a branch in the low hybrid range of frequency (LHRF) with $f\sim f_{LH}\sim\sqrt{f_{ci}f_{ce}}$ when $k_\perp$ is large. In addition, the lowest frequency branch ($\sim10-100{\rm MHz}$) is at ion cyclotron range of frequency (ICRF) with $f\lesssim f_{ci}$, which is called the ion cyclotron slow wave (ICW) or Alfv\'en wave branch. For small $k_\perp$,  the branch with the frequency slight above the slow ICW mode is called the ion cyclotron fast wave, which is actually the same branch as the LHW mode branch. Since this branch has the frequency ranging from $f<f_{ci}$ to $f\sim f_{LH}$, is also known as the high harmonic fast wave (HHFW) or helicon wave when $f\sim 10-50f_{ci}$, as well as the whistler wave\cite{Pinsker2015}. Hence, the cold plasma waves are recognized as three ranges of frequencies, i.e., ECRF, LHRF, and ICRF. Specialized to the studying of low frequency Alfv\'en wave ($<10{\rm MHz}$), people also mention it as the fourth range of frequency, to distinguish from ICRF.

Secondly, the magenta dash lines show all the six branches of warm multi-fluid plasma waves (here, $S=2$, two-fluid), with  $T_{s0}=1{\rm keV}$. By comparing with the cold plasma waves, we see two differences: a new low frequency ($\sim2{\rm MHz}$) branch, and the modification of the previous cold branches by thermal effects at large $k_\perp$. For the former, the new branch (a zero frequency branch is omitted) is usually called as ion acoustic wave (IAW) with $\omega\sim k_\parallel v_{ts}$, which vanishes for $T_{s0}=0$. For the latter, we see that the low hybrid $f\to f_{LH}$ and upper hybrid $f\to f_{UH}=\sqrt{f_{pe}^2+f_{ce}^2}$ resonances at $k_\perp\to\infty$ do not exist any more, but be continually modified into the new thermal branches. The ion cyclotron slow wave is also modified from $f<f_{ci}$ to $f>f_{ci}$ at large $k_\perp$.

Thirdly, the solid lines show the kinetic branches. Two new features are manifested: (a) many branches exist and it is difficult to obtain all the solutions to them; (b) the frequency $f$ becomes complex number, and ${\rm Im}(f)$ represents the wave damping. It is interesting to noted that the kinetic solutions deviate from the cold plasma solutions only when the warm multi-fluid solutions also deviate from the cold plasma solutions, which implies that the warm multi-fluid model is sufficient to tell whether a cold plasma model is valid for given waves. However, the kinetic solutions are much different from the warm multi-fluid solutions when the thermal effects are non-ignorable. For IAW and ICW slow modes, they are strongly damped and more than one sub-branche of them can be found. In Fig.\ref{fig:pinsker15_fig7}, we only show some of the kinetic IAWs and ICWs, but much more can be found in the kinetic dispersion relation, which will be discussed in Appendix \ref{sec:howmanymodes}. Most of them are strongly damped and difficult to be excited and thus are less physically interesting. For LHW, we see that it is separated into many different branches at the high order ion cyclotron harmonic frequencies. There exists at least one branch at each $lf_{ci}<{\rm Re}(f)<(l+1)f_{ci}$, with $l$ being a positive integer number, i.e., LHW is modified by the ion Bernstein wave (IBW). It is also interesting to find that the envelop curve of these LHWs or IBWs are close to the curve of the multi-fluid solution. For ECW, we also see the modification into cyclotron harmonics, i.e., the electron Bernstein waves (EBW). The slow X (SX) mode changes continually to an EBW, which is the SX-B mode conversion. It is also found that although the multi-fluid slow-X branch deviates from both cold solutions and kinetic solutions at large $k_\perp$, it is close to the envelop of the peak values of the kinetic EBWs.

From the above discussions, we conclude that the thermal effects in multi-fluid model change or mitigate the cold plasma UH, LH and IC resonances. The thermal effects in kinetic model further leads to Bernstein modes when $k_\perp\rho_{cs}\gtrsim1$ and collisionless/Landau damping when $v_{ts}\gtrsim\omega/k_\parallel$. For small $k_\perp$ modes and high frequency ECW X and O modes, the cold plasma model is well valid. The  asymptote of warm fluid curve can be in agreement with the envelop curve of kinetic model, though the kinetic model breaks the fluid model to multi-branche Bernstein modes. These mean that the warm fluid model can capture the main physics of the thermal effects, though it can not capture the separation effects from kinetic Bernstein modes.

\section{Model equations}\label{sec:model}

The model equations to obtain Fig.\ref{fig:pinsker15_fig7} and the following waves accessibility diagrams are presented in this section. Here, we limit our discussions to linear plasma waves in an infinite uniform system, and assume the background magnetic field ${\bm B}_0=(0,0,B_0)$, wave vector ${\bm k}=(k_x,0,k_z)=(k\sin\theta,0,k\cos\theta)$, i.e., $k_\parallel=k_z$, $k_\perp=k_x$. We have $S$ species with index $s=1, 2, \cdots, S$, and omit the drift velocities of each species, i.e., ${\bm v}_{s0}=0$. The electric charge, mass, density and temperature of each species are $q_s$,  $m_s$, $n_{s0}$ and $T_{s0}$, respectively. We discuss three models, i.e., cold plasma model, multi-fluid warm plasma model, and kinetic model.  The MHD model is usually limited to only low frequency waves ($f<f_{ci}$) and thus are not discussed here.

\subsection{Cold plasma}\label{sec:cold}
The plasma waves in (multi-fluid) cold plasma model have been well discussed in textbooks and monographes \cite{Stix1992,Swanson2003,Gurnett2005}. Here we only summarize some main results and equations.
Using Stix notations \cite{Stix1992}, we have the following relation for perturbed electric field ${\bm E}=(E_x,E_y,E_z)^T$,
\begin{eqnarray}\label{eq:dreq}
{\bm D}\cdot{\bm E}={\bm n}\times({\bm n}\times{\bm E})+{\bm K}\cdot{\bm E}=0,
\end{eqnarray}
with
\begin{eqnarray}\nonumber
&  {\bm K} =\left( {\begin{array}{ccc}
  S & -iD & 0\\
  -iD & S & 0\\
  0 & 0 & P\end{array}}\right),
\end{eqnarray}
and
\begin{eqnarray*}
S=1-\sum_s\frac{\omega_{ps}^2}{\omega^2-\omega_{cs}^2},~~P=1-\sum_s\frac{\omega_{ps}^2}{\omega^2},\\D=\sum_s\frac{\omega_{cs}}{\omega}\frac{\omega_{ps}^2}{\omega^2-\omega_{cs}^2},\\
R=1-\sum_s\frac{\omega_{ps}^2}{\omega(\omega+\omega_{cs})},~~L=1-\sum_s\frac{\omega_{ps}^2}{\omega(\omega-\omega_{cs})},\\
{\bm n}=\frac{{\bm k}c}{\omega},~\omega_{cs}=\frac{q_sB_0}{m_s},~\omega_{ps}=\sqrt{\frac{n_{s0}q_s^2}{\epsilon_0m_s}},~~c=\frac{1}{\sqrt{\mu_0\epsilon_0}},
\end{eqnarray*}
where $\omega_{ps}$ and $\omega_{cs}$ are the plasma frequency and cyclotron frequency of each species (note that for electron, $q_e<0$ and thus $\omega_{ce}<0$), $c$ is the speed of light, $\epsilon_0$ is the permittivity of free space, $\mu_0$ is the permeability of free space, and $\bm n$ is the refractive index vector.
Since ${\bm E}\neq0$, we obtain the cold plasma dispersion relation
\begin{eqnarray}\label{eq:colddr}
\bar{D}({\bm n},\omega)=|{\bm D}|=An^4-Bn^2+RLP=0,
\end{eqnarray}
where $A=S\sin^2\theta+P\cos^2\theta$ and $B=RL\sin^2\theta+PS(1+\cos^2\theta)$, which  gives
\begin{eqnarray}
n^2=\frac{B\pm F}{2A},
\end{eqnarray}
where $F$ can be written in a positive definite form
\begin{eqnarray}
F^2=(RL-PS)^2\sin^4\theta+4P^2D^2\cos^2\theta,
\end{eqnarray}
which means $n^2=n_\parallel^2+n_\perp^2$ should always be real number for real $\tan\theta=n_\perp/n_\parallel$. After solving Eq.(\ref{eq:colddr}) to obtain the dispersion tensor $\bm D$, we can use Eq.(\ref{eq:dreq}) to obtain the polarizations of the wave perturbed electric field $\bm E$. Eq.(\ref{eq:colddr}) is a quadratic equation for $n^2$, which can readily be solved and is sufficient for us to obtain the cold plasma wave accessibility diagrams for all wave modes $n^2$. Obviously, only four cold plasma modes of $n$ exist for a given frequency $\omega$. Due to the symmetry of the positive and negative direction, only two of them need to be considered. However, to answer how many $\omega$ exist for a given $\bm k$, we need some other forms of cold plasma model equations, which will be discussed in subsection \ref{sec:fluid}. The simple answer is that $3S+4$ cold plasma waves of $\omega$ exist. Due to the symmetry, only half of them need to be considered. That is, for two species ($S=2$), we have $(3\times2+4)/2=5$ branches as shown in Fig.\ref{fig:pinsker15_fig7}.

\subsection{Warm multi-fluid plasma}\label{sec:fluid}
We start with the mulit-fluid equations\cite{Swanson2003,Xie2014}
\begin{subequations} \label{eq:mfeq0}
\begin{eqnarray}
  & \partial_t n_s = -\nabla\cdot(n_s\bm v_s),\\
  & \partial_t \bm v_s = -\bm v_s\cdot \nabla\bm v_s+\frac{q_s}{m_s}(\bm E+\bm v_s\times \bm B)-\frac{\nabla P_s}{\rho_s},\\
  & \partial_t \bm E = c^2\nabla\times\bm B - \bm J/\epsilon_0,\\
  & \partial_t \bm B = -\nabla\times\bm E,
\end{eqnarray}
\end{subequations}
where the current and pressure equations are
\begin{subequations} \label{eq:fpeq2}
\begin{eqnarray}
  & \bm J = \sum_sq_sn_s\bm v_s, \\  
  & d_t(P_s\rho_s^{-\gamma_s}) = 0,
\end{eqnarray}
\end{subequations}
with the mass density $\rho_{s}\equiv m_sn_{s}$.  According to Ref.\cite{Xie2019}, to be close to the results of kinetic model, we set the adiabatic coefficient $\gamma_s=2$ instead of $\gamma_s=5/3$.

Linearizing the equations using $f=f_0+\delta f$ and $\delta f=\delta fe^{i({\bm k}\cdot{\bm r}-\omega t)}$, we obtain
\begin{subequations} 
\begin{eqnarray}
  &\delta \bm J = \sum_sq_sn_s\bm v_s, \\  
  & \delta P_s=P_{s0}\gamma_s\delta\rho_s=c_s^2m_s\delta n_s,
\end{eqnarray}
\end{subequations}
where $c_s^2\equiv \gamma_sP_{s0}/\rho_{s0}=\gamma_sk_BT_{s0}/m_s$ and $P_{s0}=n_{s0}k_BT_{s0}$, and thus
\begin{eqnarray}
\nabla\delta P_s=(ik_x\delta P_s,0,ik_z\delta P_s).
\end{eqnarray}

The linearized equations of Eqs.(\ref{eq:mfeq0}) are
\begin{subequations} \label{eq:mfeq1}
\begin{eqnarray}
 & \omega\delta n_s = n_{s0}(k_x\delta v_{sx}+k_z\delta v_{sz}),\\
  & \omega\delta v_{sx}=i\frac{q_s}{m_s}(\delta E_x+\delta v_{sy}B_0)+\frac{k_xc_s^2\delta n_s}{n_{s0}},\\
  & \omega\delta v_{sy}=i\frac{q_s}{m_s}(\delta E_y-\delta v_{sx}B_0),\\
  & \omega\delta v_{sz}=i\frac{q_s}{m_s}\delta E_z+\frac{k_zc_s^2\delta n_s}{n_{s0}},\\
& \omega\delta E_x=-c^2(-k_z\delta B_y)-i\frac{\sum_sq_sn_{s0}\delta v_{sx}}{\epsilon_0},\\
& \omega\delta E_y=-c^2(k_z\delta B_x-k_x\delta B_z)-i\frac{\sum_sq_sn_{s0}\delta v_{sy}}{\epsilon_0},\\
& \omega\delta E_z=-c^2(k_x\delta B_y)-i\frac{\sum_sq_sn_{s0}\delta v_{sz}}{\epsilon_0},\\
& \omega\delta B_x=-k_z\delta E_y,\\
& \omega\delta B_y=k_z\delta E_x-k_x\delta E_z,\\
& \omega\delta B_z=k_x\delta E_y.
  \end{eqnarray}
\end{subequations}
We have $4S+6$ equations and thus have $4S+6$ solutions for $\omega$ when other parameters are fixed (Note: Two of the solutions could be $\omega=0$).  All the numerical solutions of $\omega$ can be easily obtained by the eigenvalue solver\cite{Xie2014,Xie2019}. In the cold plasma limit, $c_s^2=0$, the above equations can reduce to $3S+6$ equations, which yield $3S+6$ solutions for $\omega$, with two of them being $\omega=0$.

However, in this tutorial, we are interested in the solutions of $\bm k$, especially $k_x$, for a given $\omega$ and other parameters. We may consider two cases: (1) fixed $\theta$ to solve $k$, and (2) fixed $k_z$ to solve $k_x$. We only discuss the later case in the present tutorial, since there are usually some constraints of the $k_\parallel$ spectrum due to the antenna. Eqs.(\ref{eq:mfeq1}) can be rewritten as
{\scriptsize
\begin{subequations} \label{eq:mfeq1kx}
\begin{eqnarray}
  k_xc_s^2 \delta n_s &=& \omega n_{s0}\delta v_{sx}-\frac{n_{s0}q_s}{\omega m_s}\Big[ik_zc^2\delta B_y+\\&&\frac{\sum_{s'}q_{s'} n_{s'0}\delta v_{s'x}}{\epsilon_0}-\omega_{cs}(\delta E_y-\delta v_{sx}B_0)\Big],\\\nonumber
  k_xn_{s0}\delta v_{sx} &=& \omega\delta n_s-\Big(i\frac{q_s}{m_s}\frac{k_zn_{s0}}{\omega}\delta E_z+\frac{k_z^2c_s^2}{\omega}\delta n_s\Big),\\
  k_x\delta E_y &=& \omega\delta B_z,\\
  k_x\delta E_z &=& \Big(\frac{k_z^2c^2}{\omega}-\omega\Big)\delta B_y-i\frac{k_z}{\omega}\frac{\sum_{s'}q_{s'} n_{s'0}\delta v_{s'x}}{\epsilon_0},\\
  k_xc^2\delta B_y &=& -\omega\delta E_z+\frac{1}{\omega}\sum_{s'}\Big(\omega_{ps'}^2\delta E_z-i\frac{k_zc_{s'}^2q_{s'}}{\epsilon_0}\delta n_{s'}\Big),\\
  k_xc^2\delta B_z &=& \Big(\omega-\frac{k_z^2c^2}{\omega}\Big)\delta E_y-\sum_{s'}\frac{\omega_{ps'}^2}{\omega}(\delta E_y-\delta v_{s'x}B_0),
  \end{eqnarray}
\end{subequations}}
where we have $2S+4$ equations for $k_x$ and thus have $2S+4$ solutions of $k_x$ for a given other parameters. Note also that in the first equation, $\frac{n_{s0}q_s}{m_s}\frac{\sum_{s'}q_{s'}\nonumber n_{s'0}\delta v_{s'x}}{\epsilon_0}\neq\frac{\sum_{s'}q_{s'}^2\nonumber n_{s'0}^2\delta v_{s'x}}{m_{s'}\epsilon_0}$.  In the cold plasma limit, $c_s^2=0$, the above equations can be reduced to 4 equations and thus have 4 branches of $k_x$ in the cold plasma, which agrees with the  standard case in Sec.\ref{sec:cold}.

Eqs.(\ref{eq:mfeq1kx}) can readily yield a matrix eigenvalue equation
\begin{eqnarray}
  k_x\cdot{\bm X}={\bm M}\cdot{\bm X},
\end{eqnarray}
with ${\bm X}=[\delta n_s,\delta v_{sx},\delta E_y,\delta E_z,\delta B_y, \delta B_z]^{T}$. The numerical solutions of all the eigenvalues $k_x$ and corresponding eigenvectors $\bm X$ can be easily obtained by a standard eigen matrix solver. The eigenvector $\bm X$ represents the polarizations of the wave. 

Use the above model, we can obtain all the wave solutions of multi-fluid plasmas with temperature effects. Hence, this provides a simple and fast way to know whether a cold plasma wave propagation model is valid for given parameters.

\subsection{Kinetic model}\label{sec:kinetic}
We only consider Maxwellian velocity distribution function for each species
\begin{eqnarray}
  f_{s0}=\frac{1}{\pi^{3/2}v_{ts}^3}e^{-\frac{v_\parallel^2+v_\perp^2}{v_{ts}^2}},
\end{eqnarray}
with thermal velocity $v_{ts}=\sqrt{\frac{2k_BT_{s0}}{m_s}}$, where $k_B$ is the Boltzmann constant. Note that some authors may use $v_{ts}=\sqrt{\frac{k_BT_{s0}}{m_s}}$ and thus the following dispersion relation would be slightly different. The dispersion relation is
\begin{eqnarray}\nonumber
  \bar{D}(\omega,{\bm k})&=&|{\bm D}(\omega,{\bm k})|\\
  &=&|{\bm K}(\omega,{\bm k})+({\bm k}{\bm k}-k^2{\bm I})\frac{c^2}{\omega^2}|=0,
\end{eqnarray}
i.e.,
\begin{eqnarray}\label{eq:drkinetic}\nonumber
&  \bar{D}(\omega,{\bm k}) =\left| {\begin{array}{ccc}
  D_{xx} & D_{xy} & D_{xz}\\
  D_{yx} & D_{yy} & D_{yz}\\
  D_{zx} & D_{zy} & D_{zz}\end{array}}\right|=\\&\left| {\begin{array}{ccc}
  K_{xx}-\frac{k_z^2c^2}{\omega^2} & K_{xy} & K_{xz}+\frac{k_zk_xc^2}{\omega^2}\\
  K_{yx} & K_{yy}-\frac{k^2c^2}{\omega^2} & K_{yz}\\
  K_{zx}-\frac{k_zk_xc^2}{\omega^2} & K_{zy} & K_{zz}-\frac{k_x^2c^2}{\omega^2}\end{array}}\right|=0,
\end{eqnarray}
where
\begin{eqnarray}
  {\bm K}={\bm I}+{\bm Q}={\bm I}-\frac{{\bm \sigma}}{i\omega\epsilon_0},~~{\bm Q}=-\frac{{\bm \sigma}}{i\omega\epsilon_0}.
\end{eqnarray}
Define 
\begin{eqnarray*}
  a_s=k_\perp\rho_{cs}, ~~b_s=a_s^2=k_\perp^2\rho_{cs}^2,~~\rho_{cs}=\sqrt{\frac{k_BT_s}{m_s}}\frac{1}{\omega_{cs}}=\frac{v_{ts}}{\sqrt{2}\omega_{cs}},
\end{eqnarray*}
and
\begin{eqnarray*}
  \zeta_{sn}=\frac{\omega-n\omega_{cs}}{k_zv_{ts}}.
\end{eqnarray*}
Note for electron $q_s<0$, and thus $\omega_{cs},\rho_{cs},a_s<0$. After standard derivations\cite{Stix1992,Swanson2003,Xie2019}, we obtain
\begin{eqnarray}
  {\bm K}={\bm I}+\sum_s\frac{\omega_{ps}^2}{\omega^2}\Big[\sum_{n=-\infty}^{\infty}\zeta_{s0}Z(\zeta_{sn}){\bm X}_{sn}+2\zeta_{s0}^2{\bm L}\Big],
\end{eqnarray}
and
\begin{eqnarray}\nonumber
  {\bm X}_{sn} =\left( {\begin{array}{ccc}
  \frac{n^2}{b_s}\Gamma_{sn} & in\Gamma'_{sn} & \sqrt{2}\zeta_{sn}\frac{n}{a_s}\Gamma_{sn}\\
  -in\Gamma'_{sn} & \frac{n^2\Gamma_{sn}}{b_s}-2b_s\Gamma'_{sn} & -i\sqrt{2}\zeta_{sn}a_s\Gamma'_{sn}\\
  \sqrt{2}\zeta_{sn}\frac{n}{a_s}\Gamma_{sn} & i\sqrt{2}\zeta_{sn}a_s\Gamma'_{sn} & 2\zeta_{sn}^2\Gamma_{sn} \end{array}}\right),
\end{eqnarray}
where $Z(\zeta)=\frac{1}{\sqrt{\pi}}\int_{-\infty}^{+\infty}\frac{e^{-z^2}}{z-\zeta}dz$ is the plasma dispersion function\cite{Xie2013}, $\Gamma_{sn}\equiv\Gamma_n(b_s)$ and  $\Gamma'_{sn}\equiv\Gamma'_n(b_s)$, with
\begin{eqnarray*}
  \Gamma_n(b)=I_n(b)e^{-b},~~\Gamma'_n(b)=(I'_n-I_n)e^{-b},\\
  I'_n(b)=\frac{I_{n+1}+I_{n-1}}{2},~~I_{-n}=I_{n},\\
  \Gamma_{sn}^{''}(b)=\Big(\frac{I'_{n+1}+I'_{n-1}}{2}+I_n-2I'_n\Big)e^{-b},\\
  Z'(\zeta)=-2[1+\zeta Z(\zeta)],
\end{eqnarray*}
and $I_n$ is the $n$-th order modified Bessel function. In $\bm L$, only $L_{zz}=1$, and other elements vanish. A spectral method\cite{Weideman1995,Xie2013} is provided to fast and accurate calculation of $Z(\zeta)$ function integral with analytic continuation, which is used in the present tutorial. The dispersion relation can be written out terms by terms as 
\begin{eqnarray}\nonumber
 & \bar{D}(\omega,{\bm k})=D_{xx}D_{yy}D_{zz}+D_{yx}D_{zy}D_{xz}+D_{zx}D_{yz}D_{xy}-\\&D_{xz}D_{yy}D_{zx}-D_{yz}D_{zy}D_{xx}-D_{zz}D_{yx}D_{xy}=0.
\end{eqnarray}
Thus the calculating of the derivatives of $D(\omega,{\bm k})$, e.g.,  $\partial D(\omega,{\bm k})/\partial\omega$, $\partial D(\omega,{\bm k})/\partial k_x$ and $\partial D(\omega,{\bm k})/\partial k_z$, is also straightforward though tedious. These derivatives would be useful for root finding and ray tracing calculations.

Due to the complication of the plasma dispersion function $Z(\zeta)$ and the infinite order summation of the Bessel functions, it is difficult to know how many waves exists in the above kinetic system. A matrix method is developed in Ref.\cite{Xie2016,Xie2019} to obtain all the important $\omega$ solutions of the above kinetic model based on Pad\'e approximation for $Z$ function. The kinetic solutions in Fig.\ref{fig:pinsker15_fig7} are obtained use this matrix method and be further confirmed in kinetic root finding based on accurate $Z$ function. To obtain all solutions of $k$ or $k_\perp$ is more challenging, for which however, we can still use Cauthy contour approach to know how many solutions in a complex plane region (Appendix \ref{sec:howmanymodes}). After we know the rough location of a mode, we can use some standard root finding approach, such as Newton iterative $x^{n+1}=x^n-f(x_n)/f'(x_n)$,  with the initial guess to solve the dispersion relation Eq.(\ref{eq:drkinetic}) to obtain the kinetic solution.


\begin{widetext}

\begin{figure}[htbp]
\centering
\includegraphics[width=14cm]{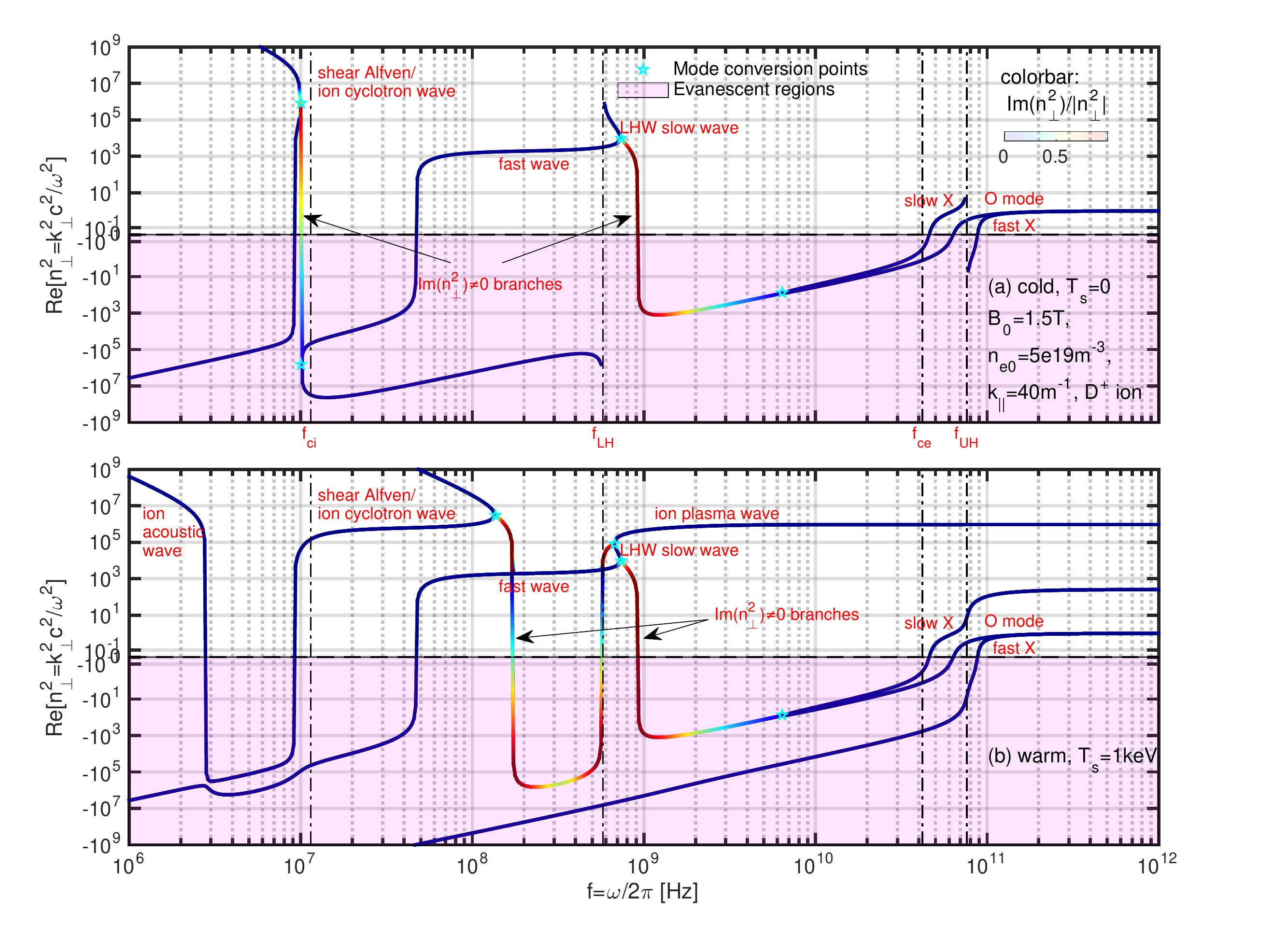}
\caption{Comparison of the cold and fluid warm plasma waves, with the same parameters as in Fig.\ref{fig:pinsker15_fig7}.}\label{fig:cmp_bon_nper2}
\end{figure}

\end{widetext}


\section{Accessibility Diagrams}\label{sec:accessibility}
As mentioned in Sec.\ref{sec:intro}, we can use $n_\perp^2$ to access the plasma wave accessibility. Hence, for the accessibility diagrams, we mean the diagrams of $n_\perp^2$ v.s. other parameters, such as frequency, density and so on. Since $n_\perp^2$ varies from very small number to very large number, and can be both positive and negative, it requires to rescale the plot. Instead of the $\log_{10}(n_\perp^2)$ scale, which is not suitable for negative numbers, we use the $\sinh^{-1}(an_\perp^2)$ scale, as suggested in Ref.\cite{Pinsker2015}, with $a$ being a rescale factor to zoom in the small $|n_\perp^2|$ regions. We set $a=10$ as default.

The theoretical aspects of the plasma wave accessibility, resonance and cutoff, for ECW, LHW and ICW, based on the cold plasma theory can be found in textbooks\cite{Wesson2011,Freidberg2007} and we do not repeat too much here. Some of the kinetic thermal effects are also shown in Ref.\cite{Cairns1991}. We mainly focus on the numerical diagrams in the present tutorial and use these diagrams to illustrate the physics behind. 

Before discussing the waves in each range of frequencies, we firstly take a look at Fig.\ref{fig:pinsker15_fig7}, which is recalculated and presented in Fig.\ref{fig:cmp_bon_nper2}, where we fix the frequency $f$ to solve the $k_\perp$ and then obtain $n_\perp^2$ for cold plasma and warm multi-fluid solutions. This is essentially the same as Fig.\ref{fig:pinsker15_fig7}. Comparing to Fig.\ref{fig:pinsker15_fig7}, we can see two differences in Fig.\ref{fig:cmp_bon_nper2}: (1) the waves in evanescent regions with $n_\perp^2<0$ are shown; (2) we can see clearly how different branches are connected to each other with the mode conversion points being shown. For examples, the ECRF O mode is connected to slow X mode in the evanescent regions, which means that in some conditions the O mode can convert to slow X mode, i.e., the usual called O-X conversion. By further considering the kinetic result in Fig.\ref{fig:pinsker15_fig7}, the slow X mode is connected to EBW, i.e., which gives a full story of the O-X-B mode conversion. 
The modes with non vanish ${\rm Im}(n_\perp^2)$ in Fig.\ref{fig:cmp_bon_nper2} do not exist in Fig.\ref{fig:pinsker15_fig7}, which is because that the $\tan\theta=k_\parallel/k_\perp$ of these modes in Fig.\ref{fig:cmp_bon_nper2} are not real numbers. In Fig.\ref{fig:pinsker15_fig7}, we have assumed both $k_\parallel$ and $k_\perp$  to be real. Thus, the accessibility diagram Fig.\ref{fig:cmp_bon_nper2} can not only show whether a wave can propagate, but also reveal the connections between each wave modes.

\begin{widetext}

\begin{figure}[htbp]
\centering
\includegraphics[width=12cm]{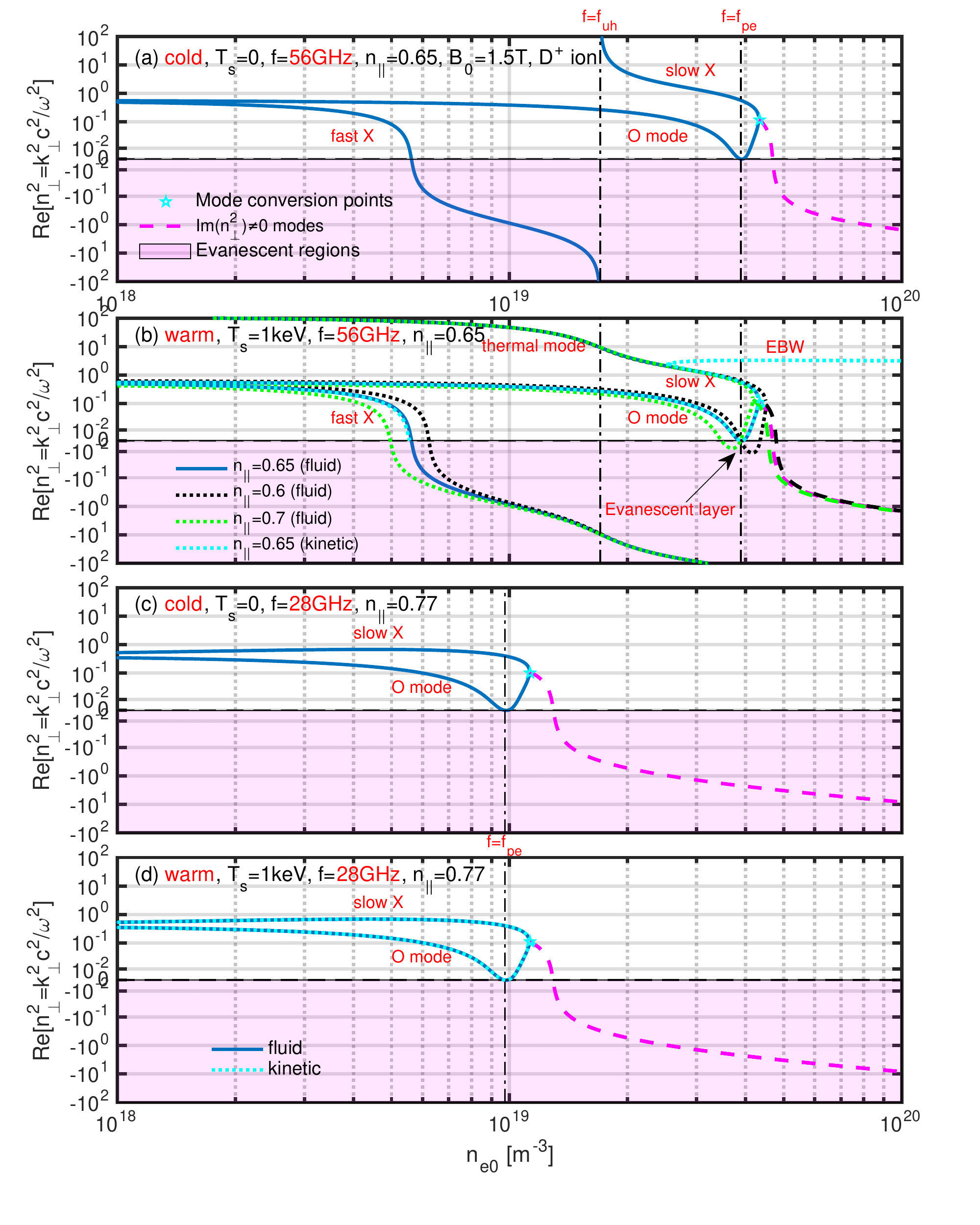}
\caption{Comparison of the cold and fluid warm plasma ECW waves for $f=28$ and $56$GHz, O-SX-B and O-SX as a function of density $n_{e0}$.}\label{fig:ecw_bon_scann_nper2}
\end{figure}

\end{widetext}


\subsection{ECRF}\label{sec:ecrf}
For the two branches vacuum ECW O-mode and X-mode, $n^2\lesssim1$, thus we set $n_\parallel<1$ to plot the accessibility diagram of this range of frequency. A good review of ECRF and EBW heating can be found in Ref.\cite{Laqua2007}. Ref.\cite{Ali2013} also provides an intuitive introduction of different ways for the mode conversion to EBW.

Fig.\ref{fig:ecw_bon_scann_nper2} shows two typical frequencies with $f=28{\rm GHz}<f_{ce}=42{\rm GHz}$ and $f=56{\rm GHz}>f_{ce}$ for magnetic field $B_0=1.5{\rm T}$. For the parameters in Fig.\ref{fig:ecw_bon_scann_nper2}, we see that no mode can propagate in high density range, i.e., both O-mode and X-mode have a certain cutoff density. This is why ECRF can not be used to heating high density plasmas beyond the cutoff density,  unless we use even higher wave frequencies. By choosing an optimized $n_\parallel$ in Fig.\ref{fig:ecw_bon_scann_nper2}(b), we can minimize the width of the evanescent layer, which facilitates the O-SX mode conversion. In the cold plasma model, the SX mode meets the UH resonant layer and can not pass that layer. In the warm fluid model, the SX mode can convert into a thermal mode, which is relevant to EBWs in the  kinetic model. This diagram demonstrates how O-X-B mode conversion happens. For Fig.\ref{fig:ecw_bon_scann_nper2}(c) and (d), we see that the thermal effects are ignorable, where the O mode can also convert to SX mode. Fig.\ref{fig:ecw_bon_scann_nper2}(a) and (b) can represent low field side launch; whereas Fig.\ref{fig:ecw_bon_scann_nper2}(c) and (d) can represent high field side launch. When the ECRF can mode conversion  to EBW via O-SX-B (O-X-B), it can overcome the cutoff density and heat the high density plasma. However, if only the FX-SX-B (X-B) mode conversion happens, ECRF would mainly still be limited to heat low density plasma.


\begin{widetext}

\begin{figure}[htbp]
\centering
\includegraphics[width=12cm]{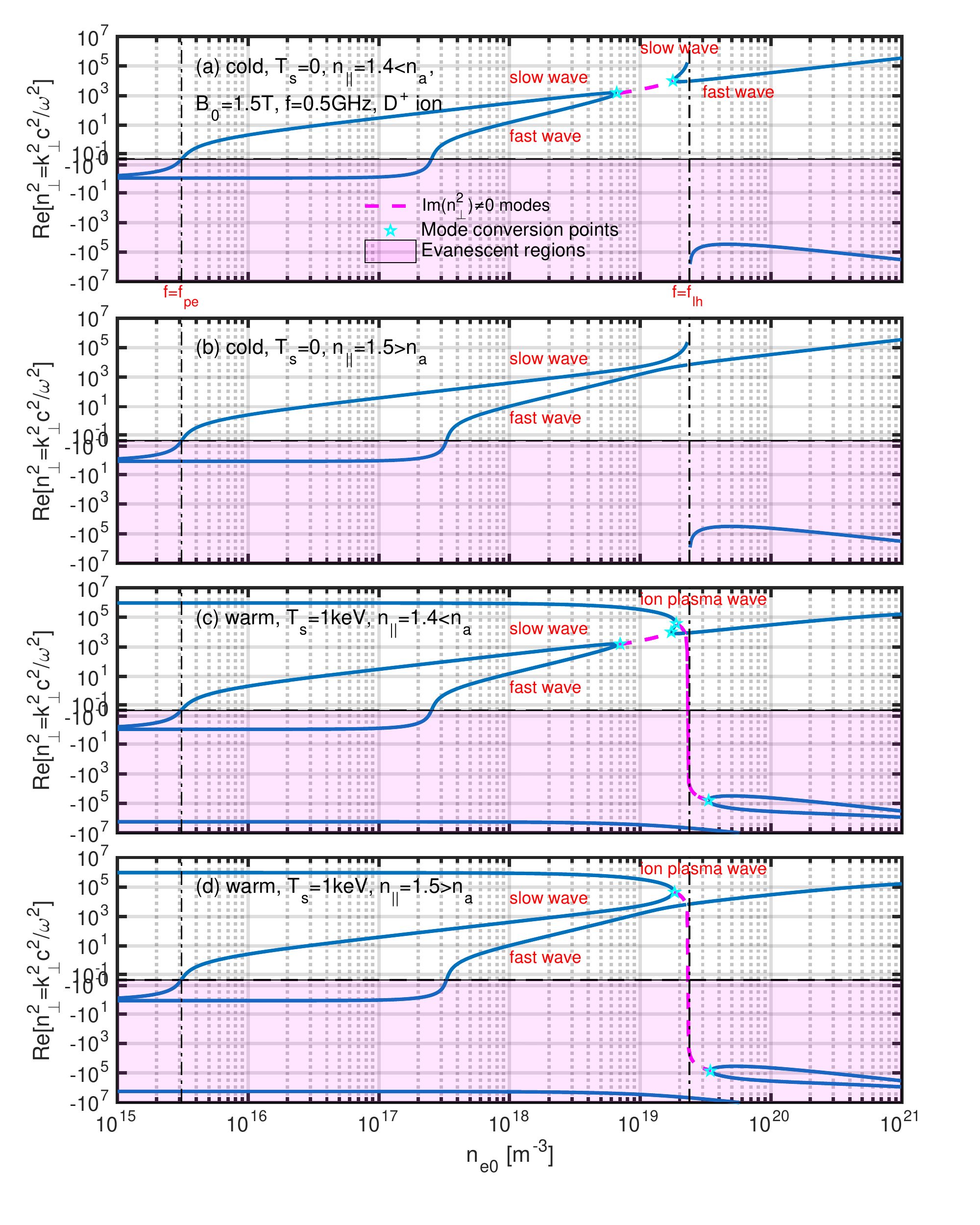}
\caption{Comparison of the cold and fluid warm plasma LHW waves for $f=500$MHz, with the same parameters as in Fig.\ref{fig:pinsker15_fig7} versus density.}\label{fig:lhw_bon_scann_nper2}
\end{figure}

\begin{figure}[htbp]
\centering
\includegraphics[width=12cm]{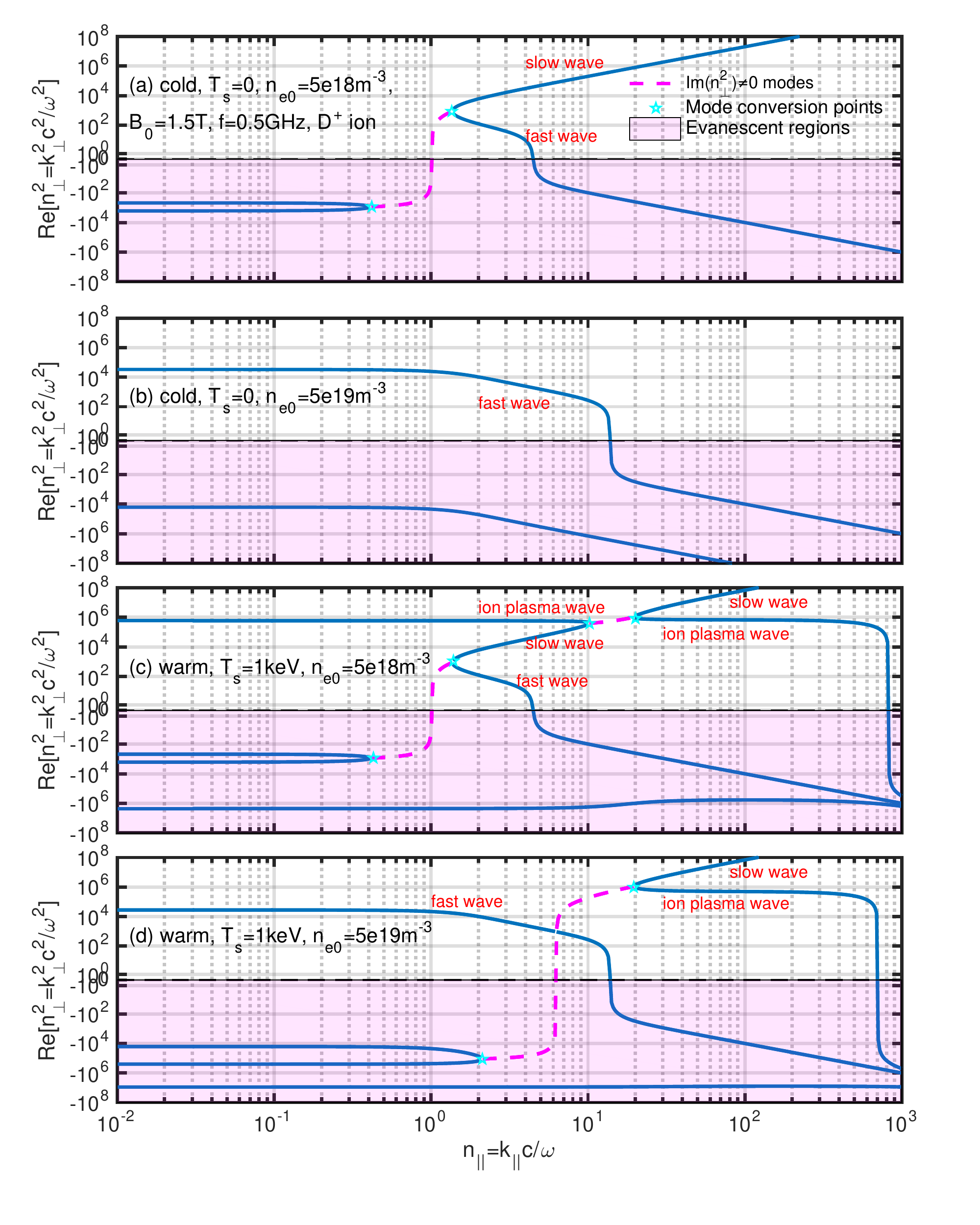}
\caption{Comparison of the cold and fluid warm plasma LHW waves for $f=500$MHz, with the same parameters as in Fig.\ref{fig:pinsker15_fig7}, versus $n_\parallel$, for densities $n_{e0}=5\times10^{18}$ and $5\times10^{19}$m$^{-3}$. For even lower densities, only the ECW with $n_\parallel<1$ branches exist.}\label{fig:lhw_bon_scannpar_nper2}
\end{figure}

\end{widetext}


\subsection{LHRF}\label{sec:lhrf}
The accessibility of LHW was well described in literature\cite{Stix1965,Bonoli1982,Pinsker2015,Wesson2011}. We use Fig.\ref{fig:lhw_bon_scann_nper2} and Fig.\ref{fig:lhw_bon_scannpar_nper2} to show the main physics. There exist two cold plasma wave modes in this range of frequency, the LHW fast wave (FW) and slow wave (SW). The SW is launched by the antenna at the low density edge and propagate to the core plasma. There exists a critical parallel refractive index $n_{\parallel}=n_a$, with $n_a=\sqrt{S}+\sqrt{-D^2/P}\simeq\sqrt{1/(1-\omega^2/|\omega_{ci}\omega_{ce}|)}$ in the cold plasma model\cite{Freidberg2007,Pinsker2015}. If the wave with small $n_{\parallel}<n_a$, the SW will convert to FW and reflect back; whereas if $n_{\parallel}>n_a$, the SW will propagate to the LH resonant layer and be strongly absorbed. This is why we require a slightly high $n_\parallel$ for LHW to meet the accessibility requirement. For the warm plasma, the situation is slightly different, i.e., the wave will not stop at the LH resonant layer but further convert to a thermal ion plasma wave. Due to this reason, LHW was proposed to heating ions in early years\cite{Bonoli1982,Wesson2011}. However, this conversion to thermal ion plasma wave is usually not effective. Later, people found that LHW can be much efficient to heat electron and drive current\cite{Fisch1987}, due to the electron thermal velocity $v_{te}$ being close to the wave phase velocity $v_p=\omega/k_\parallel$. The best condition is considered as $v_{te}<v_p<3v_{te}$. The Landau damping is smaller for smaller $v_p$; whereas for larger $v_p$, less particles are involved in the wave-particle interaction, thus also leading to a smaller heating and current driven efficiency. Though LHW has a very high current driven efficiency,  the major drawback is that it is difficult to propagate to the high density core plasma. We can see from Fig.\ref{fig:lhw_bon_scannpar_nper2} that both FW and SW can propagate in the low density ($n_{e0}=5\times10^{18}{\rm m}^{-3}$) case; whereas only FW can propagate in the high density case ($n_{e0}=5\times10^{19}{\rm m}^{-3}$).


\begin{widetext}

\begin{figure}[htbp]
\centering
\includegraphics[width=12cm]{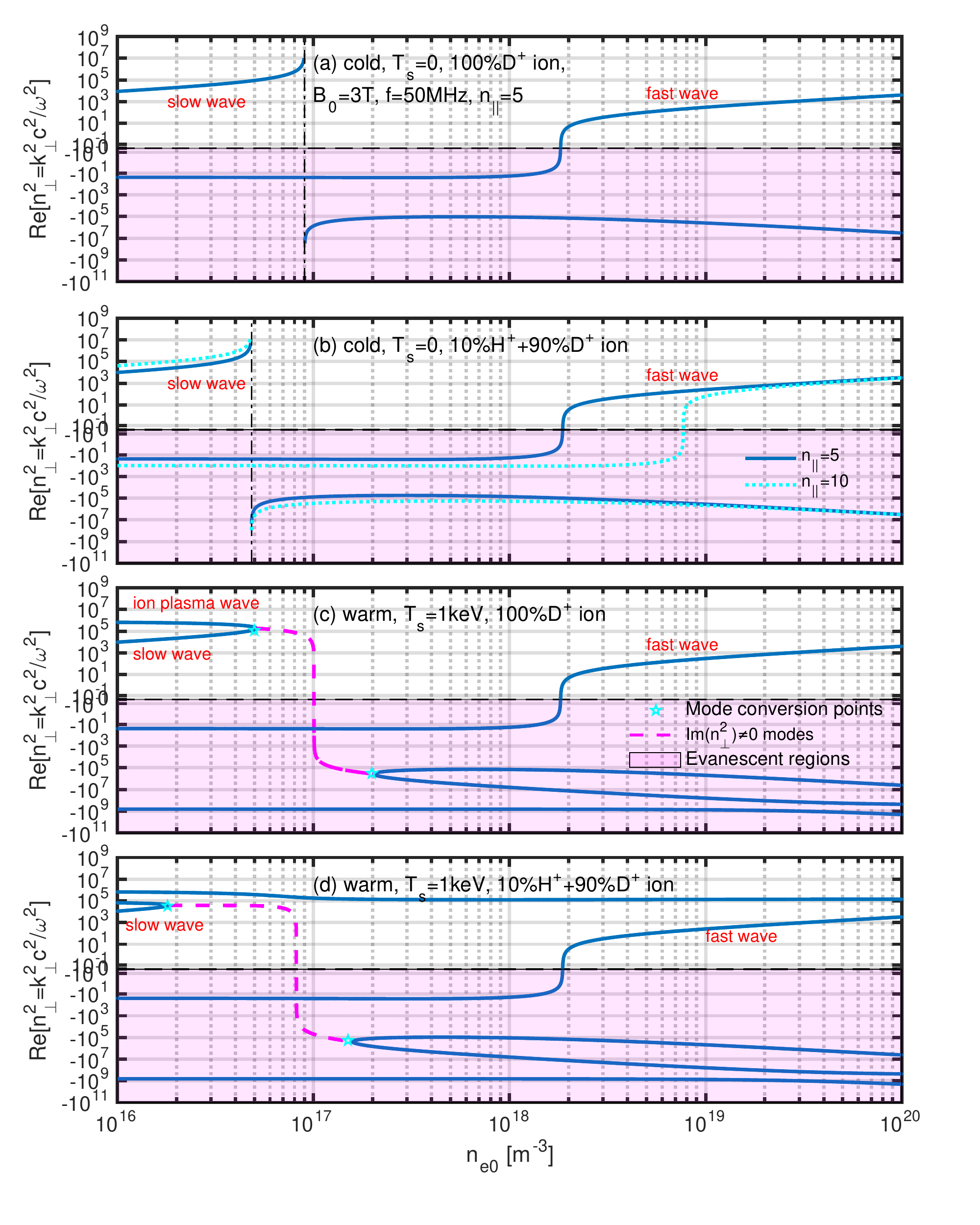}
\caption{Comparison of the cold and fluid warm plasma ICW waves for $f=50$MHz, FW and SW  versus density $n_{e0}$.}\label{fig:icw_bon_scann_nper2}
\end{figure}

\begin{figure}[htbp]
\centering
\includegraphics[width=14cm]{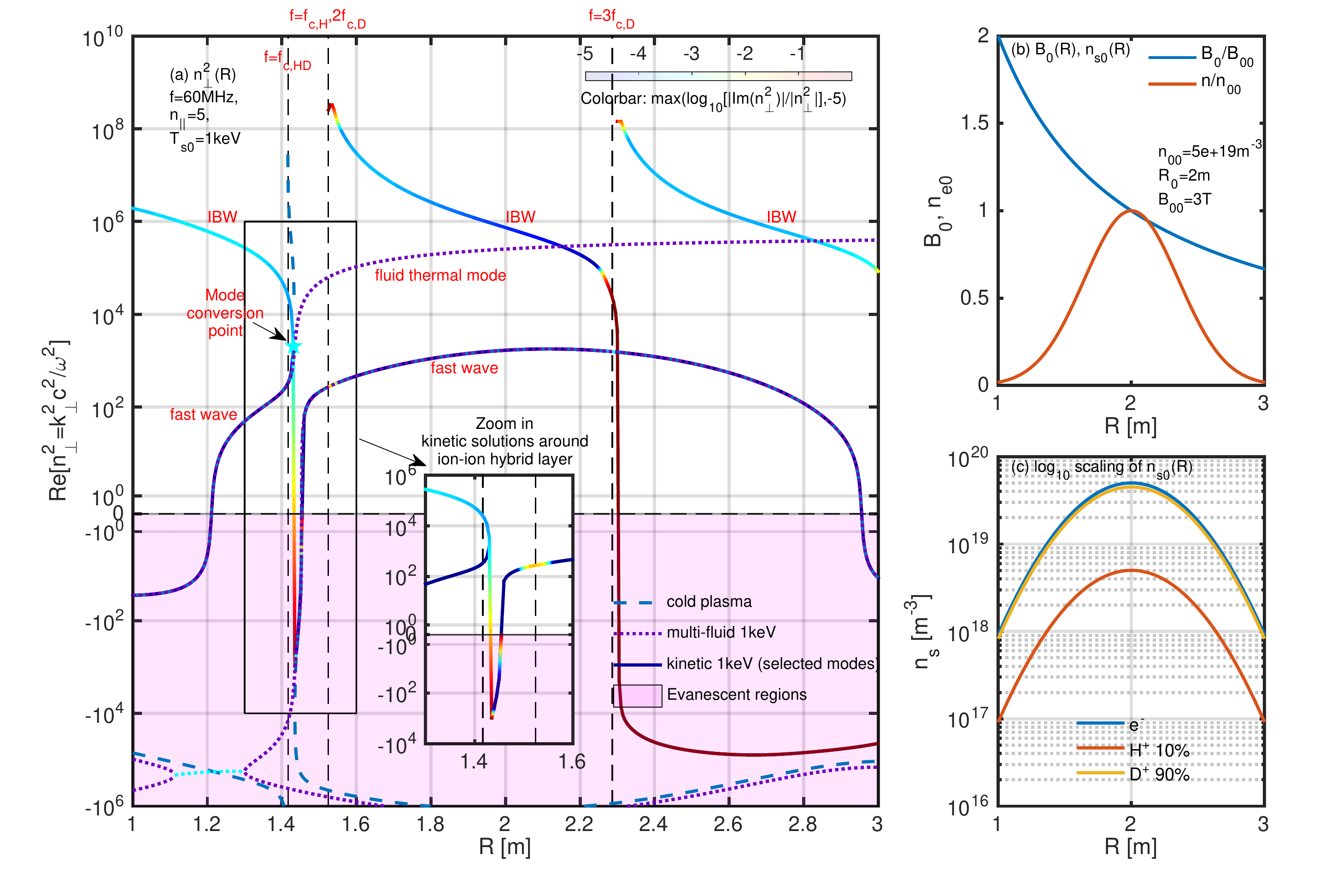}
\caption{Comparison of the cold, fluid warm and kinetic ICW waves for $f=60$MHz, FW and SW versus $R$ for different $n_{s0}(R)$ and $B_0(R)$.}\label{fig:icw_scanR_nper2}
\end{figure}

\end{widetext}


\subsection{ICRF}\label{sec:icrf}

There are also two wave modes present in the ICW range of frequencies, the ICW fast wave (the same branch of LHW) and ICW slow wave (shear Alfv\'en wave).  Fig.\ref{fig:icw_bon_scann_nper2} shows the typical accessibility of ICRF. We see that the SW can only propagate in a much lower density range; whereas the FW can only propagate in a much higher density range. The thermal effects are significant to SW but almost invisible to FW. Due to the low frequency and long wave length of ICW SW, instead of the present local accessibility analysis, a global model is usually required to describe it. Here, we mainly discuss the ICW FW.
We see in Fig.\ref{fig:icw_bon_scann_nper2}(b), larger $n_\parallel$ (comparing $n_\parallel=10$ and $n_\parallel=5$) requires larger density for FW propagation, which implies that we need small $n_\parallel$ to meet the accessibility condition at the low density plasma edge. 

The fundamental ion cyclotron resonant heating of ions are thought to be inefficient for ICRF. Three heating approaches are proposed: (a) second order or high order ion cyclotron resonant heating; (b) multi-ions with minor ions hybrid resonance heating; (c) IBW heating. Compare to the one ion species case ($100\% ~{\rm D}^+$), we see that in Fig.\ref{fig:icw_bon_scann_nper2}, the two ion species case ($10\%~ {\rm H}^++90\% ~{\rm D}^+$) mainly affects the SW.

To demonstrate the above mentioned three heating approaches of ICRF, we choose a two ion species case ($10\%~ {\rm H}^++90\% ~{\rm D}^+$) and plot the accessibility as $n_\perp^2$ versus the major radius $R$ in Fig.\ref{fig:icw_scanR_nper2}, which may represent a realistic ICRF heating scenario. The magnetic field is taken as $B(R)=B_{00}R_0/R$ with $R_0=2{\rm m}$ and $B_{00}=3{\rm T}$. The density profiles of each species are $n_{s0}=n_{s00}\exp[-(R-R_0)^2/\Delta R^2]$ with $\Delta R=0.5{\rm m}$ and  $n_{s00}=[1.0,0.1,0.9]\times n_{00}$ for electron, H and D ions, with $n_{00}=5\times10^{19}{\rm m}^{-3}$. The temperature is taken to be constant for simplicity, with $T_{s0}(R)=1{\rm keV}$. We see in Fig.\ref{fig:icw_scanR_nper2} that the FW is separated by the ion-ion hybrid resonant layer into two regions in $R$. In both regions, the FW solutions from the cold plasma model, warm fluid mode and kinetic model are similar. When thermal effects are important at the resonant position, the FW converts to a thermal mode in the fluid model, but to IBW in the kinetic model. Kinetic damping is obvious at the ion-ion resonant position $\omega=\omega_{ii}=[\omega_{c1}\omega_{c2}(n_1m_1+n_2m_2)]/(n_1m_2q_1/q_2+n_2m_1q_2/q_1)$,  and the cyclotron resonant positions $\omega=l\omega_{ci}$ with $l=2,3$, and also for  the IBWs. In conclusion, the cold plasma model is usually sufficient to study the accessibility and wave trajectory in the ray tracing of ICRF FW, but not valid for SW and IBWs. The mode conversion at the high field side from FW to IBW will be shown in Sec.\ref{sec:ray} using the kinetic ray tracing.

\begin{widetext}

\begin{figure}[htbp]
\centering
\includegraphics[width=14cm]{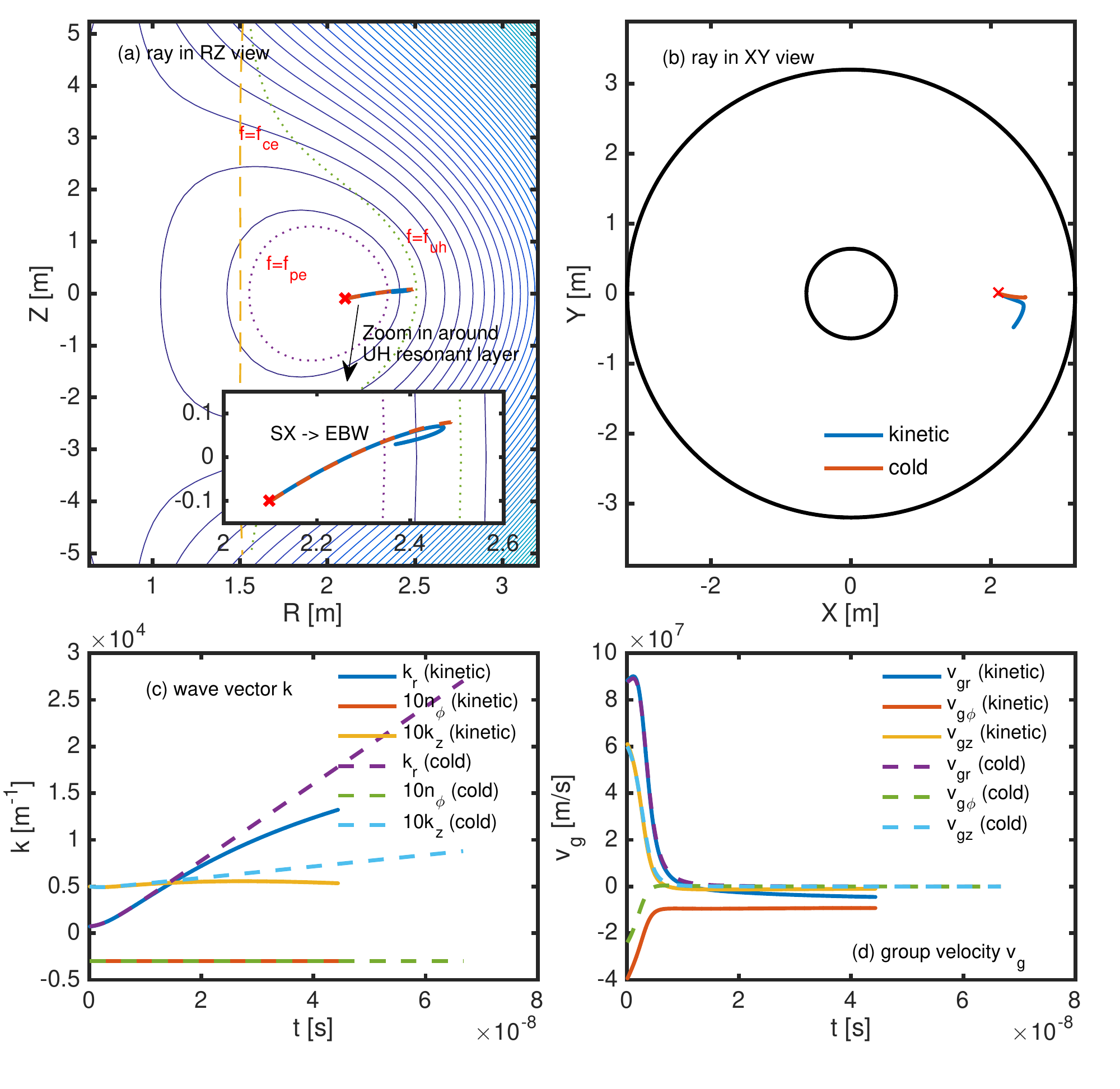}
\caption{Comparison of the cold and kinetic ECW ray tracings for $f=56$GHz. Mode conversion from SX to EBW is shown in the kinetic model; the ray stops at the UH layer in the cold model.}\label{fig:ecw_ray} 
\end{figure}

\begin{figure}[htbp]
\centering
\includegraphics[width=14cm]{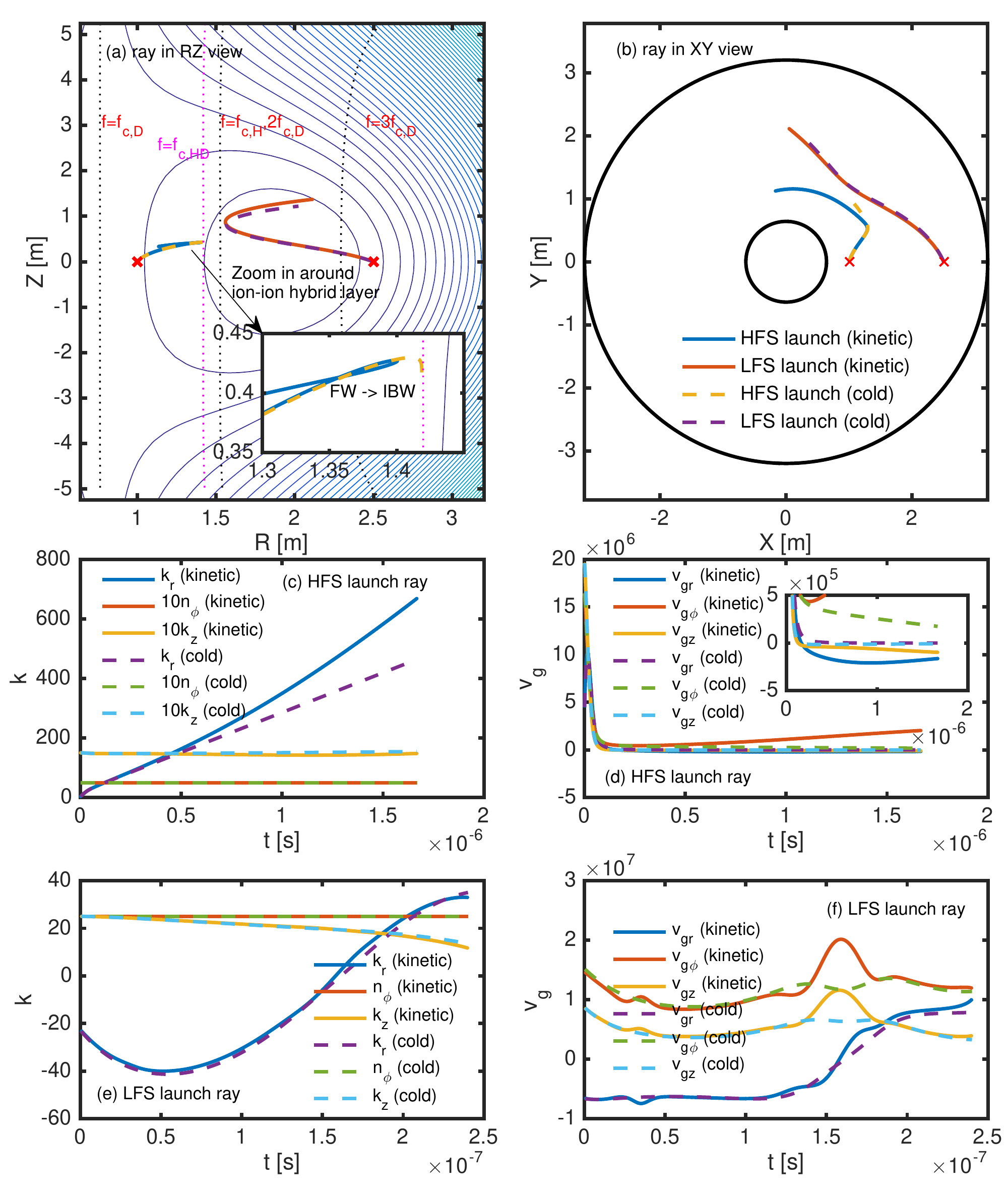}
\caption{Comparison of the cold and kinetic ICRF FW ray tracings for $f=60$MHz. In both high field side and low field side launch cases, the ray trajectories are similar for FW under cold and kinetic models. However, for the HFS launch, the FW may convert to IBW at the ion-ion hybrid layer under kinetic model, but stop ($v_{gr}\to0$)  at this layer under the cold plasma model.}\label{fig:icw_ray} 
\end{figure}

\end{widetext}


\section{Ray tracing}\label{sec:ray}
In this section, we also discuss some ray tracing results to understand how valid the cold plasma model is for the ray trajectory. There existed several ray tracing codes based on kinetic model for plasma waves heating in 1970s, e.g., for the study of ICRF\cite{McVey1979} and ECRF\cite{Maekawa1978}. However, most later ray tracing codes\cite{Batchelor1982,Smirnov2003,Peysson2012,Xie2021} choose the cold plasma model for ray trajectory as default option. The main reasons are that the kinetic calculation is much slower than the calculation in the cold model, and the cold model is found to be valid for most cases. Here, we revisit this problem to help understand the validation range of the cold plasma model for ray tracing. 

Performing the coordinates transformation from $(x,y,z,k_x,k_y,k_z)$ to $(r,\phi,z,k_r,n_\phi,k_z)$, we can have the ray tracing equations\cite{Xie2021}
\begin{eqnarray}\label{eq:dray}
\frac{dr}{d\tau}=\frac{\partial D}{\partial k_r},~
\frac{d\phi}{d\tau}=\frac{\partial D}{\partial n_\phi},~
\frac{dz}{d\tau}=\frac{\partial D}{\partial k_z},~\\
\frac{dk_r}{d\tau}=-\frac{\partial D}{\partial r},~
\frac{dn_\phi}{d\tau}=-\frac{\partial D}{\partial\phi},~
\frac{dk_z}{d\tau}=-\frac{\partial D}{\partial z},~
\end{eqnarray}
with
\begin{eqnarray}\label{eq:dtdtau}
\frac{dt}{d\tau}=-\frac{\partial D}{\partial \omega},
\end{eqnarray}
Usually, the dispersion relation is written as $D=D(\omega,k_\parallel,k_\perp)=0$.
Here, the parallel wave vector $k_\parallel={\bm k}\cdot{\bm b}=\frac{1}{B}\Big(k_rB_r+k_zB_z+\frac{n_\phi}{r}B_\phi\Big)$ is defined for the magnetic field $\bm B$, with $k_\perp^2=k^2-k_\parallel^2$, $B=B(r,z)=\sqrt{B_r^2+B_z^2+B_\phi^2}$, $k^2=k_r^2+k_z^2+\frac{n_\phi^2}{r^2}$. We use the analytical Solov\'ev equilibrium to demonstrate the results, with $R_0=2{\rm m}$, $R_x=0.8{\rm m}$, $q_0=3$, $E=1.5$, $\tau=0.8$, $L_{ns}=0.9$, $L_{Ts}=0.8$ and $n_{e0}=5\times10^{19}{\rm m}^{-3}$. One can refer to BORAY\cite{Xie2021} code for details of the ray tracing method and equilibrium model used here. The cold plasma dispersion relation Eq.(\ref{eq:colddr}) and kinetic Eq.(\ref{eq:drkinetic}) are substituted into the above ray tracing equations (\ref{eq:dray})-(\ref{eq:dtdtau}) to calculate the ray trajectories. 

Firstly, we show the ECW case, with $B_0=1.5{\rm T}$, D ions and $T_{s0}=0.5{\rm keV}$. We start from a slow X mode in the plasma. Fig.\ref{fig:ecw_ray} shows that for the cold plasma, the SX ray meets the UH layer and stops there with the group velocity $v_g\to0$; whereas for kinetic plasma, the SX ray converts to EBW and reflects back with $v_{gr}>0$ to $v_{gr}<0$. This is the process described in section \ref{sec:ecrf}.

We also show the ICW case to demonstrate the process described in section \ref{sec:icrf}., with $B_0=3.0{\rm T}$, $10\%~ {\rm H}^++90\% ~{\rm D}^+$ and $T_{s0}=10{\rm keV}$. We start from FW in the plasma for both high field side (HFS) launch and low field side (LFS) launch. Fig.\ref{fig:icw_ray} shows that for the cold plasma, the FW HFS ray meets the ion-ion hybrid resonant layer and stops there with the group velocity $v_g\to0$; whereas for the kinetic plasma, the FW ray converts to IBW and reflects back with $v_{gr}>0$ to $v_{gr}<0$. The LFS rays are similar for both cold and kinetic models. The computation times for the kinetic (keep the Bessel summation to order $n=5$) and cold rays with 2000 times steps are around 100s and 0.1s respectively. That is, due to the much slow speed of the kinetic calculation, it is acceptable to choose cold plasma model for ray tracing when the thermal effects are less important.

If we further integrate the damping rate ${\rm Im}(\omega)$ or ${\rm Im}(k_\perp)$ along the ray, we can obtain the absorption rate of the wave, which represents the wave heating efficiency. One can refer to the ray tracing method for more details.

\section{Summary and Conclusion}\label{sec:summ}

We discuss the plasma waves accessibility in this tutorial and show several useful diagrams for the wave frequency from high to low. The thermal effects are discussed and demonstrated by a warm multi-fluid model and a kinetic model. The fluid model is written as an eigenvalue matrix form, which is particular useful to reveal how many wave modes exist and can be numerically solved. It is interesting to note that the warm multi-fluid model, though lack of the Bernstein modes, can provide a quick way to decide whether the thermal effects are ignorable. The physics of the plasma waves accessibility are also demonstrated in the ray tracing calculations for the ECRF slow X mode to EBW and ICRF SW to IBW mode conversions. Cauthy contour integral method is used to study the number of kinetic wave modes. We conclude that the multi-fluid plasma modes and the conventional Bernstein modes are sufficient for most interesting physics scenarios. Other modes are usually strongly damped and are not interesting. The waves accessibility and ray tracing studies in this tutorial are limited to the linear local model. Global full wave models\cite{Brambilla1999} are required to study the fine structures of the mode conversion. In addition, the Fokker-Planck\cite{Harvey1992,Lin-Liu2003} collision model is required to study the current drive. For ECRF heating, the relativistic effect may also be important\cite{Mazzucato1987,Swanson2003}. Global three dimensional nonlinear kinetic simulations for plasma wave heating and mode conversion in a realistic toroidal geometry is available recently\cite{Bao2016}. Wave polarizations can also provide useful information, which are not discussed here.


\acknowledgments The authors would like to thank the discussions and comments from Banerjee Debabrata, Shao-dong Song, Hou-yang Guo, Xin-jun Zhang, Jian Bao and Bo-jiang Ding.


\begin{widetext}

\begin{figure}[htbp]
\centering
\includegraphics[width=14cm]{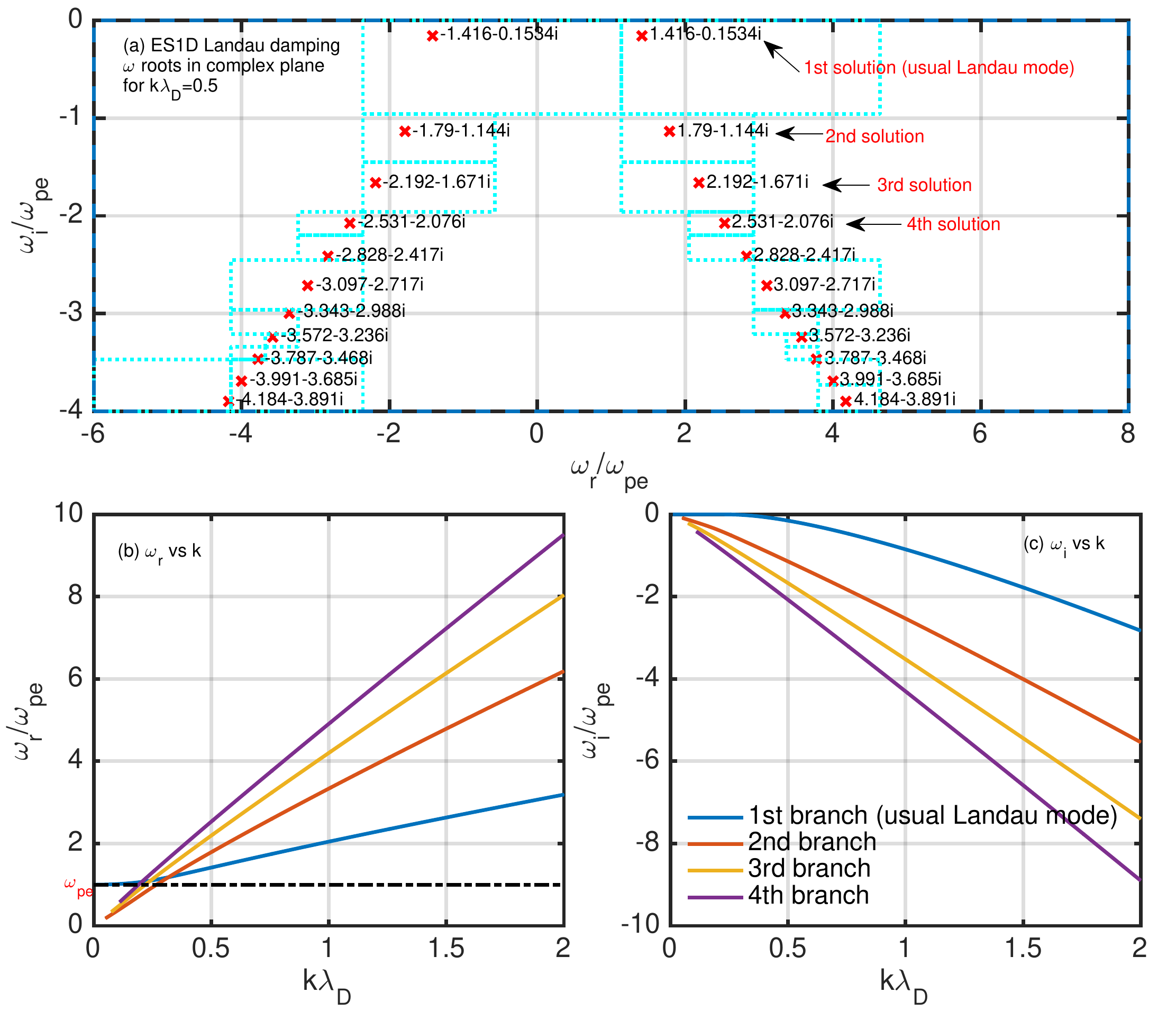}
\caption{Electrostatic 1D Landau damping roots of $\omega=\omega_r+i\omega_i$ in the complex plane for a given $k\lambda_D$. Series modes exist. In textbooks, usually only the first branch is discussed. The main reason is that other branches are strongly damped and less physically interesting.}\label{fig:es1d_landau_roots}
\end{figure}

\begin{figure}[htbp]
\centering
\includegraphics[width=14cm]{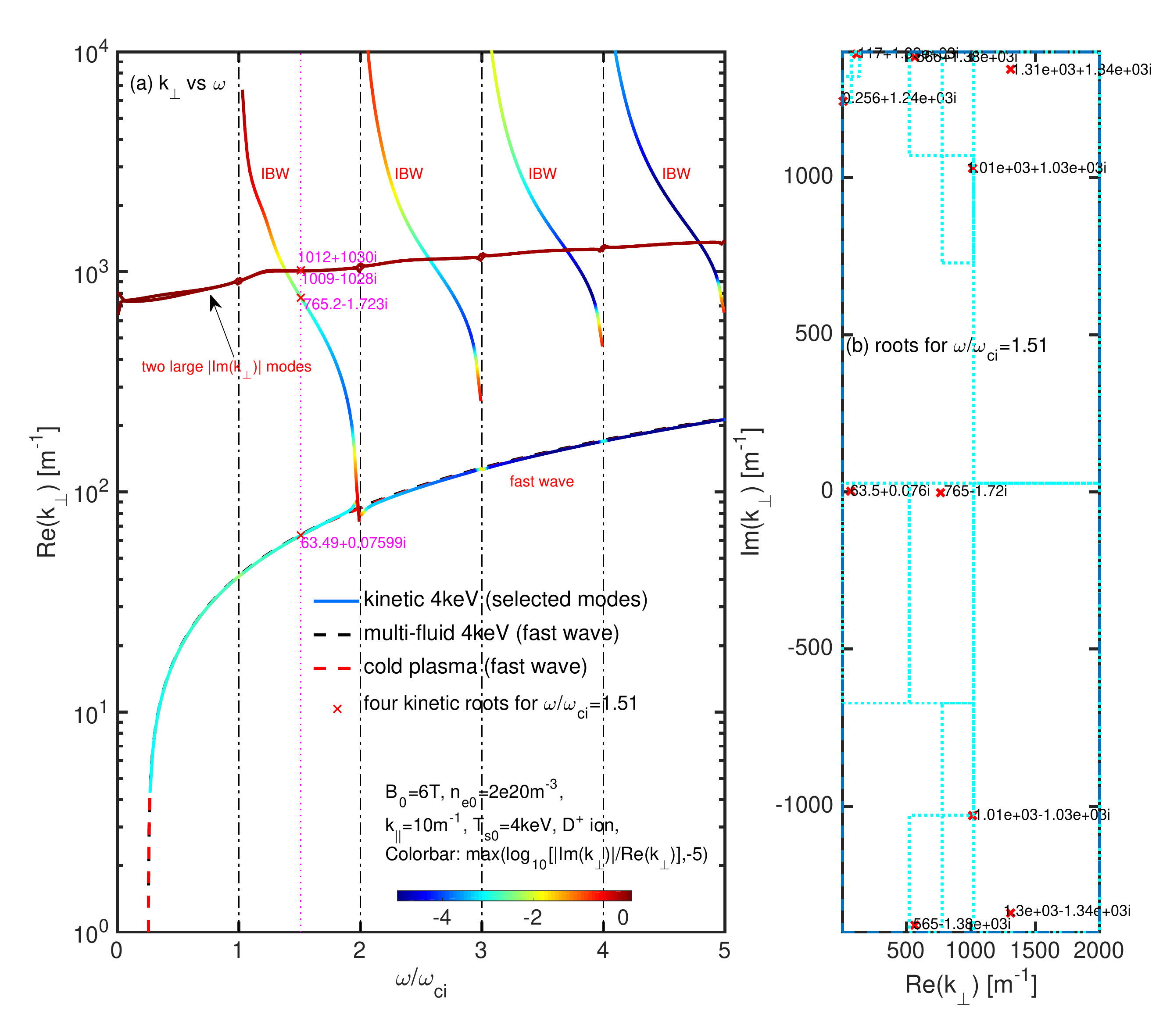}
\caption{Wave modes in the ICW range, for both FW and IBW. Both the cold and warm fluid fast wave agree with kinetic results well, which implies that the cold plasma ray tracing for fast wave is good enough. More roots exist with larger ${\rm Im}(k_\perp)$.}\label{fig:mcvey78_4keV_mi=2mp}
\end{figure}
\end{widetext}


\appendix
\section{Wave modes}\label{sec:howmanymodes}
There are two questions: (1) how many are plasma wave modes; and (2) how to find the solutions of them? For the cold plasma and warm multi-fluid plasma, the answers to these two questions are trivial based on the descriptions in Sec.\ref{sec:model}, i.e., $3S+6$ solutions for $\omega$ and $4$ solutions for $k_\perp$ in the cold plasma, and $4S+6$ solutions for $\omega$ and $2S+4$ solutions for $k_\perp$ in the multi-fluid warm plasma. All of them can be found numerically using a standard matrix eigenvalue solver. Due to the symmetry of $\omega^2$, only half of them are taken into consideration, i.e., ${\rm Re}(k_\perp)>0$ and ${\rm Re}(\omega)>0$ modes. In a beam plasma system with drift $v_{s0}\neq0$, this symmetry will be broken, i.e., the ${\rm Re}(\omega)>0$ and ${\rm Re}(\omega)<0$ modes may be different.

However, for kinetic plasmas, both of the above questions are not easy to answer. For the first question, the answer could probably be infinity solutions. The second question is even more challenging.  Fortunately, some numerical technologies can help us to locate much more solutions than the simple iterative one-by-one root finding methods such as Newton iterative. The matrix method based on Pad\'e approximation of $Z(\zeta)$ function to solve all the important (except strong damped modes, where the Pad\'e approximation is not accurate) kinetic solutions of $\omega$ modes is introduced comprehensively in the PDRK/BO code\cite{Xie2016,Xie2019}. Here, we discuss the Cauthy integral method\cite{Kravanja2000} without approximation to the dispersion relations, to reveal the distributions of the kinetic modes of $\omega$ and $k_\perp$ in the complex plane. The method is well discussed in Ref.\cite{Kravanja2000}, and the numerical code used here is taken from section 9.9 of Ref.\cite{Xie2018}. The idea behind this method is that, based on Cauthy theorem, the number of zeros $N$ of function $f(z)$ in a closed complex plane region $C$ is
\begin{eqnarray}\label{eq:nc}
 N=\frac{1}{2\pi i}\oint_C\frac{f'(z)}{f(z)}dz,
\end{eqnarray}
where $f'(z)=df/dz$. If there are high order zeros $(z-z_0)^m=0$ or poles $1/(z-z_0)^m=0$ in $f(z)$, the above conclusion would be slightly different. For the present study, we assume that we have only zeros of order $m=1$ without poles. The Nyquist plasma instability analysis method\cite{Gurnett2005} is actually based on the above Eq.(\ref{eq:nc}), as well.

We firstly consider the simplest one-dimensional electrostatic (ES1D) Maxwellian plasma with immobile ions. The dispersion relation is\cite{Gurnett2005}
\begin{eqnarray}
 1+\frac{1}{k^2\lambda_D^2}[1+\zeta Z(\zeta)]=0.
\end{eqnarray}
with $\zeta=\omega/kv_{ts}$, which is widely used in textbooks to demonstrate the Landau damping physics. The results are shown in Fig.\ref{fig:es1d_landau_roots}, where we see that series modes exist for a given $k$. In textbooks, usually only the first least damped branch is discussed. The main reason is that other branches are strongly damped and are less physical interesting. This result also reveals that the number of kinetic wave solutions could probably be infinite with even only one $Z(\zeta)$ function. If we consider the infinite summation of the Bessel functions, the kinetic wave solutions can be even more.

Now, we study the complex $k_\perp$ solutions in the kinetic dispersion relation Eq.(\ref{eq:drkinetic}). We choose the ICW range. Ref.\cite{Ignat1995} found two new hot-ion Bernstein waves with a finite $k_\parallel$. Here, we find that, there are more. Fig.\ref{fig:mcvey78_4keV_mi=2mp} shows the typical numerical solutions. Besides the two small ${\rm Im}(k_\perp)$ FW and IBW modes, there exists many other modes with large ${\rm Im}(k_\perp)$. In Fig.\ref{fig:mcvey78_4keV_mi=2mp}(b), all the solutions in the plotting region calculated by Cauthy integral method are shown. The FW branch is close to the cold plasma and warm fluid solutions. The conventional IBW is the only new kinetic mode with physically interesting. The other kinetic solutions (which come from  $Z(\zeta)$ function and Bessel functions) may not be important due to large ${\rm Im}(k_\perp)$. We should also notice that though the sign of ${\rm Im}(\omega)$ can tell whether a mode is growing (${\rm Im}(\omega)>0$) or damping (${\rm Im}(\omega)<0$), the sign of ${\rm Im}(k_\perp)$ can not alone tell whether a mode is growing or damping due to the direction of group velocity. For examples, the damped FW with $v_g>0$ and thus has ${\rm Im}(k_\perp)>0$; whereas the damped EBW can have $v_g<0$ and thus ${\rm Im}(k_\perp)<0$. In the models describe in the present tutorial, no free energies, such as beam drifts or temperature anisotropy, are considered, thus all the modes should be non-growing. Hence, all the large ${\rm Im}(k_\perp)$ modes in Fig.\ref{fig:mcvey78_4keV_mi=2mp} are considered to be strongly damped modes and are not interesting.

In summary, there exist infinity kinetic plasma wave modes due to both $Z(\zeta)$ function and infinite summation of Bessel functions. However, the multi-fluid plasma modes and the conventional Bernstein modes are sufficient for most interesting physics scenorios. Other modes are usually strongly damped and are not interesting, though it is possible that they can be excited in some special situations. It would be interesting if those special situations can be identified.


\begin{thebibliography}{99}

\bibitem{Freidberg2007} J. P. Freidberg, Plasma Physics and Fusion Energy, Cambridge University Press, Cambridge, 2007.

\bibitem{Cairns1991} R. A. Cairns, Radiofrequency Heating of Plasmas, IOP, 1991.

\bibitem{Wesson2011} J. Wesson, Tokamaks, Oxford University Press, 4th edition, 2011.


\bibitem{Stix1992} T. H. Stix, Waves in Plasmas, AIP Press, 1992.

\bibitem{Swanson2003} D.G. Swanson, Plasma Waves, second ed., IOP, 2003.

\bibitem{Gurnett2005} D.A. Gurnett and A. Bhattacharjee, Introduction to Plasma Physics: With Space and Laboratory Applications, Cambridge, 2005.


\bibitem{Batchelor1982} D. B. Batchelor and R. C. Goldfinger, RAYS: a geometrical optics code for EBT, ORNL/TM-6844, 1982.

\bibitem{Smirnov2003}  A.P. Smirnov and R.W. Harvey, The GENRAY Ray Tracing Code, 2003. https://www.compxco.com/Genray\_manual.pdf. https://github.com/compxco/genray.

\bibitem{Peysson2012} Y. Peysson, J. Decker and L. Morini, A versatile ray-tracing code for studying rf wave propagation in toroidal magnetized plasmas, Plasma Physics and Controlled Fusion, 54 (2012) 045003.

\bibitem{Mazzucato1987} E. Mazzucato, I. Fidone and G. Granata, Damping of electron cyclotron waves in dense plasmas of a compact ignition tokamak, Physics of Fluids, 30 (1987) 3745-3751.


\bibitem{Xie2021} H. S. Xie, B. Debabrata, Y. K. Bai, H. Y. Zhao and J. C. Li, BORAY: An Axisymmetric Ray Tracing Code Supports Both Closed and Open Field Lines Plasmas, arXiv:2105.12014, 2021.


\bibitem{Pinsker2015}  R. I. Pinsker, Whistlers, helicons, and lower hybrid waves: The physics of radio frequency wave propagation and absorption for current drive via Landau damping, Physics of Plasmas 22, 090901 (2015).


\bibitem{Xie2014} H. S. Xie, PDRF: A general dispersion relation solver for magnetized multi-fluid plasma, Comput. Phys. Comm. 185 (2014) 670-675.


\bibitem{Xie2019} H.S. Xie, BO: A unified tool for plasma waves and instabilities analysis, Comput. Phys. Comm. 244 (2019) 343-371. Xie, H. S., Denton, R., Zhao, J. S. and Liu, W, BO 2.0: Plasma Wave and Instability Analysis with Enhanced Polarization Calculations 
arXiv:2103.16014, 2021. https://github.com/hsxie/bo/.

\bibitem{Xie2016}  H.S. Xie, Y. Xiao, PDRK: A General Kinetic Dispersion Relation Solver for Magnetized Plasma, Plasma Sci. Technol. 18 (2) (2016) 97, http://dx.doi.org/10.1088/1009-0630/18/2/01, Update/bugs fixed at http://hsxie.me/codes/pdrk/ or
https://github.com/hsxie/pdrk/.

\bibitem{Laqua2007} H. P. Laqua, Plasma Phys. Control. Fusion 49 (2007) R1–R42.


\bibitem{Ali2013} M. Ali Asgarian, J. P. Verboncoeur, A. Parvazian, and R. Trines, Phys. Plasmas 20, 102516 (2013).

\bibitem{Bonoli1982} P. T. Bonoli and E. Ott, Toroidal and scattering effects on lower-hybrid wave propagation, Physics of Fluids 25, 359 (1982).

\bibitem{Stix1965} T. H. Stix, Radiation and Absorption Via Mode Conversion in an Inhomogeneous Collision-Free Plasma, Phys. Rev. Lett., 1965, 15, 878-882.

\bibitem{Fisch1987} N. J. Fisch, Rev. Mod. Phys. 59, 175 (1987)




\bibitem{McVey1979}   B.D. McVey, A ray-tracing analysis of fast-wave heating of tokamaks, Nuclear Fusion, 19  (1979) 461.


\bibitem{Maekawa1978}  T. Maekawa, S. Tanaka, Y. Terumichi and Y. Hamada, Wave Trajectory and Electron-Cyclotron Heating in Toroidal Plasmas, Phys. Rev. Lett., 1978, 40, 1379-1383.

\bibitem{Brambilla1999}  M. Brambilla, Plasma Phys. Control. Fusion 41 (1999) 1–34.

\bibitem{Lin-Liu2003}  Y. R. Lin-Liu, V. S. Chan, and R. Prater, Phys. Plasmas 10, 4064 (2003).

\bibitem{Harvey1992}  R.W. Harvey and M.G. McCoy, The CQL3D Code, Proc. IAEA TCM on Advances in Sim. and Modeling of Thermonuclear Plasmas, pp. 489-526, Montreal, (1992), available through USDOC/NTIS No. DE93002962; see also, CQL3D Manual, with corrections (PDF, to 2015/01/22, https://www.compxco.com/cql3d\_manual.pdf)

\bibitem{Bao2016} J. Bao, Z. Lin, A. Kuley and Z.X. Wang, Nucl. Fusion 56 (2016) 066007.

\bibitem{Kravanja2000} Peter Kravanja, Marc Van Barel, Computing the Zeros of Analytic Functions, Springer, 2000.

\bibitem{Xie2018} H. S. Xie, Introduction to Computational Plasma Physics (in Chinese), Science Press, Beijing, 2018.

\bibitem{Xie2013} H. S. Xie, Generalized plasma dispersion function: One-solve-all treatment, visualizations, and application to Landau damping, Phys. Plasmas 20, 092125 (2013).

\bibitem{Weideman1995} J. A. C. Weideman, Math. Comput. 64, 745 (1995).

\bibitem{Ignat1995} D. W. Ignat, and M. Ono, Physics of Plasmas 2, 1899 (1995).

\end{thebibliography}
\end{document}